\documentclass{aa}  
\usepackage{graphicx}
\usepackage{txfonts}

\usepackage{natbib}
\usepackage[colorlinks=true, citecolor=blue, linkcolor=blue, urlcolor=blue]{hyperref}
\usepackage{graphicx}
\usepackage{subcaption}
\usepackage{multirow}

\usepackage{dblfloatfix}

\usepackage[dvipsnames]{xcolor}

\begin{document} 

   \title{The assembly of dusty galaxies at $z \geq  4$: grain growth and the evolutionary origin of the dust-to-stellar-mass scatter}

   \author{
    Alessia D'Angelo\inst{1,2,3,4} \fnmsep\thanks{Corresponding author: alessia.dangelo@inaf.it} 
    \and Raffaella Schneider\inst{1,3,4,5} 
    \and Luca Graziani\inst{1,3,4} 
    \and Claudia Di Cesare\inst{6} 
    \and Rosa Valiante\inst{1,4} 
    \and Gerg\H{o} Popping\inst{7} 
    \and Giulia Rodighiero\inst{8,9} 
    \and Laura Bisigello\inst{9} 
    \and Denis Burgarella\inst{10} 
    \and Giovanni Gandolfi\inst{1} 
    \and Koki Otaki\inst{11}
    \and Tommaso Zana\inst{1,3,4} 
    }

   \institute{INAF/Osservatorio Astronomico di Roma, Via Frascati 33, 00078 Monte Porzio Catone, Italy 
         \and
             Dipartimento di Fisica, Università di Roma Tor Vergata, Via della Ricerca Scientifica, 1, Roma 00133, Italy
        \and
             Dipartimento di Fisica, "Sapienza" Università di Roma, Piazzale Aldo Moro 2, 00185 Roma, Italy
        \and 
             INFN, Sezione Roma1, Dipartimento di Fisica, “Sapienza” Università di Roma, Piazzale Aldo Moro 2, 00185, Roma, Italy
        \and
             Sapienza School for Advanced Studies, Viale Regina Elena 291, 00161 Roma, Italy
        \and 
            Institute of Science and Technology Austria (ISTA), Am Campus 1, 3400 Klosterneuburg, Austria
        \and
            European Southern Observatory, Karl-Schwarzschild-Str. 2, D-85748, Garching, Germany
        \and
            Dipartimento di Fisica e Astronomia, Università di Padova, Vicolo dell’Osservatorio, 3, 35122 Padova, Italy
        \and 
            INAF, Osservatorio Astronomico di Padova, Vicolo dell’Osservatorio
            5, 35122 Padova, Italy
        \and
            Aix Marseille Université, CNRS, CNES, LAM, Marseille, France
        \and
            Information Technology Center, The University of Tokyo, 6-2-3 Kashiwanoha, Kashiwa, Chiba 277-0882, Japan
}

   \date{Received date; accepted date}

\authorrunning{D'Angelo et al.}
\titlerunning{Grain growth and the origin of the dust-to-stellar-mass scatter}
 
  \abstract
   {Understanding dust evolution is essential for interpreting the physical and observable properties of galaxies across cosmic time. While nearby  galaxies have revealed a detailed picture of dust enrichment, whether the same physical picture applies in the early Universe remains unclear. Here we explore the origin of interstellar dust in simulated galaxies at $z \sim 5$-- a redshift high enough to probe the dominant dust-enrichment mechanisms at early cosmic times, yet still providing a statistically significant sample of well resolved, simulated dusty galaxies, and close to the centre of the redshift range probed by the ALPINE survey.

   Starting from the dust-to-stellar mass relation, we select five simulated star-forming galaxies with $\mathrm{SFR> 1 \, M_{\odot} yr^{-1}}$ and $\mathrm{log(M_{\star}/M_{\odot}) > 8}$. We then focus on three low-mass systems with comparable stellar masses ($\mathrm{log(M_{\star}/M_{\odot}) \sim 8.7}$) but significantly different dust masses ($\mathrm{log(M_{dust}/M_{\odot})\sim 5.1-6.2}$), to investigate the physical origin of their diversity in dust content. 
    Our analysis combines outputs from the cosmological hydrodynamical simulation \texttt{dustyGadget}, a semi-analytical model for stellar dust production, and radiative-transfer post-processing with {\sc SKIRT}.
   We confirm that stellar dust production alone reproduces only the low-dust branch of the dust-to-stellar mass relation. Grain growth is required to recover both its characteristic S-shape and the observed scatter, in agreement with the physical picture established for nearby galaxies.
   The transition to efficient grain growth appears to occur at lower metallicity than in local galaxies, consistent with the denser interstellar medium of high-redshift systems. Differences in dust content reflect different stages of galaxy assembly. Among the three low-mass systems, progressively longer star formation histories lead to greater chemical enrichment and denser ISM conditions, naturally promoting increasingly efficient grain growth. We therefore identify the assembly stage of galaxies as the key regulator of dust growth at high redshift.
  
   Mock observations indicate that JWST multi-band imaging can reveal the structural diversity of galaxies within the scatter of the dust-to-stellar mass relation, although the reconstruction of clumpy morphologies may remain incomplete. Conversely, their dust continuum emission remains below the sensitivity limits of current ALMA continuum surveys.}
  

   \maketitle

\section{Introduction}

A detailed exploration of the first low-mass galaxies has long remained challenging, primarily due to their intrinsic faintness. 
The advent of high-sensitivity facilities has opened a new observational window, allowing the interstellar medium (ISM) to be probed with unprecedented detail even at high redshift.    
In particular, multi-wavelength observations with the James Webb Space Telescope (JWST) and Atacama Large Millimeter/submillimiter Array (ALMA) are rapidly reshaping our view of early galaxy evolution (see e.g. \citealt{Herrera_Camus_2026} and \citealt{Curti2026review} for recent reviews).  
JWST observations have revealed that the mass–metallicity relation was already established within the first billion years \citep{Nakajima_massmetallicity, Curti_massmetallicity, Chemerynska2024mass-met}, implying rapid metal production, while the large scatter around the fundamental metallicity relation suggests that many high-redshift galaxies are still far from equilibrium, undergoing stochastic gas accretion and enrichment processes (“chemical chaos”, see e.g. \citealt{Heintz2023_FMR, Curti_massmetallicity}). 

ALMA dust continuum detections in massive galaxies at $z \simeq 6 - 8$ \citep{Bouwens2022, Inami_2022} have enabled to observationally investigate the dust-to-stellar and dust-to-gas mass relations \citep{Witstok2023, Algera2024}, and -- when complemented with JWST-NIRSpec data --  even dust-to-metal scaling relations \citep{Algera2026REBELS_IFU}. 

Overall, the emerging picture suggests that massive galaxies at $z<8$ are already chemically evolved, exhibiting substantial enrichment in both metals and dust. However, the observational constraints become significantly weaker at earlier cosmic times. To date, the most distant galaxy with a robust ALMA dust continuum detection lies at $z=8.31$ \citep{Tamura2019, Tamura2023, Bakx2020, Bakx2025}, while searches for dust emission in individual galaxies and stacked samples at $z>8$ have yielded only stringent upper limits on their dust content \citep{Bakx2026}.

Interestingly, these non-detections appear somewhat at odds with recent JWST observations. Although JWST cannot directly detect dust emission, it has provided indirect evidence for the presence of dust through its imprint on the photometric and spectroscopic properties of distant galaxies. In particular, several UV-luminous systems at $z \geq 9$ exhibit signatures consistent with moderate dust attenuation, such as reddened UV continuum slopes and non-zero dust visual extinction (e.g., \citealt{Cullen2023, Carniani2024, Rodighiero2026}).

A more complete picture emerges in the local Universe, where extensive observations combined with theoretical modeling have enabled detailed studies of dust assembly (e.g, see the comprehensive study by \citealt{Galliano_anearbyperspectiveondustevolution}, and the reviews by \citealt{Galliano_theinterstellardustpropertiesofnearbygalaxies} and \citealt{sorgentidipolvere}  and references therein).

These studies converge toward a two-channel formation scenario, which produces an S-shaped trend in the dust-to-gas ratio as a function of metallicity.
The transition occurs around a critical metallicity $\mathrm{Z \sim 0.2 \, Z_{\odot}}$ (see \citealt{Galliano_theinterstellardustpropertiesofnearbygalaxies,Galliano_anearbyperspectiveondustevolution}) and separates two regimes: at low metallicity, dust is primarily supplied by evolved stars (AGB stars and supernovae, see \citealt{sorgentidipolvere}), while at higher metallicity grain growth in the ISM becomes dominant. This transition is commonly interpreted as the onset of efficient accretion of gas-phase metals onto pre-existing grains, a process ultimately regulated by the star formation history and metal enrichment of galaxies \citep{Asanoetal2013, Graziani_2020}.
While interstellar grain growth is empirically supported by decades of observations of selective elemental depletion along a variety of cosmic lines of sight, and has long been incorporated into classical ISM models to compensate for the short dust destruction timescales predicted by pioneering studies (e.g, \citealt{DraineSalpeter1979, Draine1990, Jones_Hollenbach1996}), the detailed microphysics and surface chemistry governing the accretion process remain poorly understood and actively debated \citep{Ceccarelli2018}.
This process is traditionally thought to occur preferentially in dense, collapsing molecular clouds, where conditions resemble those of star formation, but require sufficient metallicity to proceed (\citealt{Galliano_theinterstellardustpropertiesofnearbygalaxies} and references therein).
Semi-empirical models further describe grain growth efficiency as a function of gas density, temperature, metallicity, and grain size (\citealt{sorgentidipolvere}, based on \citealt{Asanoetal2013}; see also \citealt{Schneideretal2016}).

Whether a similar S-shape trend \footnote{Metallicity and stellar mass are correlated, albeit with scatter, and both trace the evolutionary stage of the ISM; in the following we will therefore discuss the S-shaped trend in dust scaling relations as a function of either quantity.}  also emerges in the dust scaling relations of remote galaxies is a question that theoretical models have already begun to address, through both hydrodynamical simulations \citep[e.g.][]{Graziani_2020, Vijayan2026} and semi-analytical models \citep[e.g.][]{Popping2017,Vijayan2019}, further analyzed in subsequent comparative studies \citep[e.g.][]{PoppingPeroux2022}. 
Intriguingly, recent observations now further support this picture: at intermediate redshifts ($z \sim 2$), ALMA emission studies down to subsolar metallicities suggest that dust scaling relations are already dominated by interstellar grain growth, consistent with galaxies having passed the critical metallicity threshold \citep{Geesink2026,Solimano2026}.

JWST measurements provide observational evidence for a similar trend in the dust-to-stellar mass relation of galaxies at higher redshifts, $z\geq 4$: for a sample spanning $4<z<11.5$, a transition has been identified at an observational critical metallicity of $\rm Z\sim 0.1 \, \rm Z_{\odot}$  (e.g., \citealt{burgarella2025ceersforgingdustgrains}). This transition is accompanied by a significant scatter at low stellar masses ($\mathrm{log(M_{\star}/M_{\odot}) \sim 8-9}$), whose physical origin remains unclear.

In this work, we address these open questions -- namely, the physical origin of the observed trend and the drivers of its scatter -- using the second-generation suite of cosmological simulations performed with \texttt{dustyGadget}, the code presented in \citet{Graziani_2020}. With its improved mass resolution, this simulation suite has been extensively analysed by \citep{Di_Cesare_2023, Venditti2023}, and has been shown to reproduce the dust masses and temperatures of galaxies at $z \ge 4$ \citep{Di_Cesare_2023, Algera2024}. These higher-resolution simulations reproduce the S-shaped trend in the dust-to-stellar mass plane previously found in the lower-resolution exploratory simulation presented in \citet{Graziani_2020}, where the model was first introduced. The persistence of this trend across simulations with different numerical resolutions strongly disfavors a resolution-driven origin, indicating instead that it reflects the physical processes implemented in the model. Interestingly, the substantial scatter characterising the transition regime also persists across redshifts, further supporting its intrinsic physical origin \citep{Di_Cesare_2023}.

Focusing on galaxies at $z\sim 4.5$, we aim to answer two key questions: {\it (i)} Does the same dust physics observed locally already operate in the early Universe, despite its vastly different physical environments and conditions? {\it (ii)} Which intrinsic galactic properties regulate dust abundance, and how do they connect to current observational capabilities?

We show that the age of the Universe does not hinder the onset of grain growth; rather, the evolutionary stages of galaxies govern their ability to build up dust in their ISM. 

The paper is organised as follows.
Section \ref{sec:method} introduces \texttt{dustyGadget}, with particular attention to the dust evolution model adopted and to the rationale behind our choice of five galaxies among the full simulated sample.
Our main results are presented in Section \ref{sec:results}, followed by their physical interpretation in Section \ref{sec:discussion}, where we compare our findings with previous works and discuss the main caveats and limitations of our approach. We summarize the key messages of this study in Section~ \ref{sec:conclusions}.

\begin{table*}
\centering
\caption{Properties of the simulated galaxies at $z = 4.5$ selected for the present study. We report the stellar mass, the total dust mass, the dust mass contributed by stars only (stardust mass), the dust-to-gas mass ratio (dustiness), the metallicity (in solar units), the host halo mass, the star formation rate, and the stardust mass fraction contributed by AGB stars.}

\resizebox{\textwidth}{!}{

\begin{tabular}{ |c||c|c|c|c|c|c|c|c|c|}
 \hline
 & $\mathrm{log(M_{\star}/M_{\odot})}$ 
 & $\mathrm{log(M_{dust}/M_{\odot})}$ 
 & $\mathrm{log(M_{stardust}/M_{\odot})}$ 
 & $\mathrm{log(M_{dust}/M_{gas})}$
 & $\mathrm{Z [Z_{\odot}]}$
 & $\mathrm{12+log(O/H)}$
 & $\mathrm{log(M_{halo}/M_{\odot})}$
 &$\mathrm{SFR [M_{\odot}/yr]}$
 &$\mathrm{f_{AGB}}$
 \\ 
 \hline
 \textbf{G818}   &8.69 & 5.078 & 4.88  &-5.34 &0.04  &7.299 &11.17  & 2.35 & 0.32\\
 \textbf{G3073} &8.72 & 5.657 & 4.98   &-4.31 &0.05 &7.396   & 10.71 & 1.74& 0.37\\
 \textbf{G1910}  &8.73 & 6.248 & 5.04  &-3.85  &0.12 &7.776  &10.91 & 1.22 & 0.47\\
 \textbf{G22} &10.08 & 7.805 & 6.35  &-3.37  &0.18  &7.944   & 11.95  & 40.11 & 0.41\\
 \textbf{G5}  &10.64& 8.598 & 6.92  &-2.73 &0.37 &8.262    &12.17 & 112.26 & 0.42\\
 \hline
\end{tabular}
} 
\label{tab:summarytable}
\end{table*}

\begin{figure*}
\centering 
\includegraphics[width=0.9\textwidth]{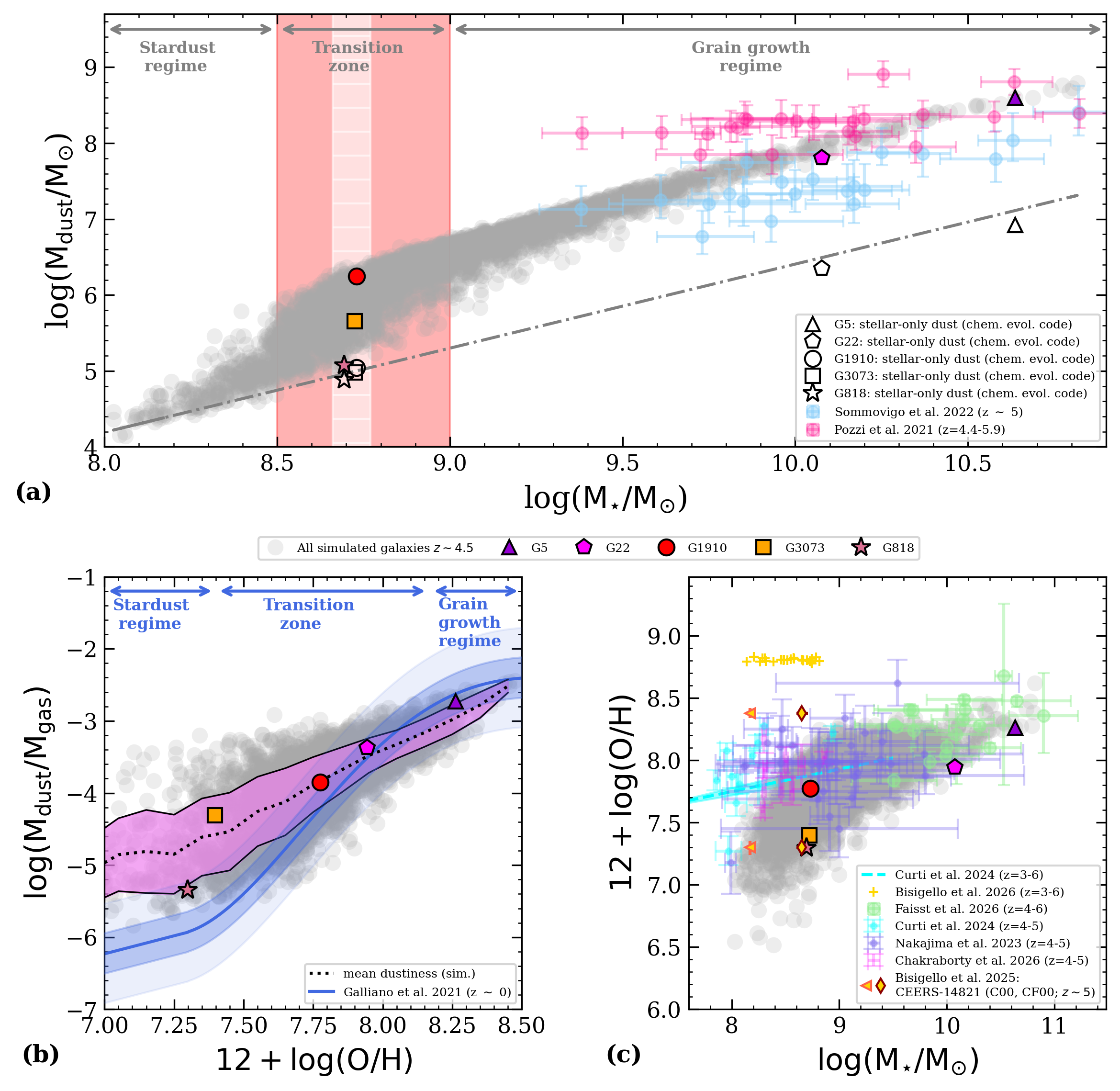}

\caption{Fundamental scaling relations at $z\sim 4.5$. Grey points show the full simulated galaxy population, while coloured symbols highlight the galaxies 2 for this study. \textbf{(a)} Dust-to-stellar mass relation. Empty symbols indicate the stellar dust component (see text). The grey dash-dotted line shows the fit to the low-mass branch. For guidance, we indicate the regimes dominated by stellar dust production, the transition region (red shaded area), and the grain-growth-dominated regime. The light pink band marks the stellar-mass range spanned by galaxies G818, G3073, and G1910. Observational measurements of star-forming galaxies at $z \sim 5$ from the ALPINE survey are shown for comparison (cyan: \citealt{Sommovigo_sshape}; pink: \citealt{Pozzi_sshape}). \textbf{(b)} Dustiness--metallicity relation. The black dotted line and violet shaded region denote the mean relation and its $1\sigma$ scatter for the simulated sample. The blue curve shows the best-fitting relation for nearby galaxies from \citet{Galliano_anearbyperspectiveondustevolution}, with the shaded areas indicating the $1\sigma$ and $2\sigma$ dispersion. \textbf{(c)} Mass--metallicity relation. Simulated galaxies are compared with recent JWST measurements at $3\lesssim z\lesssim6$, including low-mass galaxies from JADES (orange; \citealt{Curti_massmetallicity}), spectroscopically confirmed systems from the ERO, GLASS, and CEERS surveys (blue; \citealt{Nakajima_massmetallicity}), direct-$\mathrm{T_e}$ metallicity measurements from JADES and PRIMAL (green; \citealt{Chakraborty_massmetallicity}), and massive main-sequence galaxies from the ALPINE-CRISTAL-JWST survey (violet; \citealt{Faisst_massmetallicity}). We also show the HELM galaxies from CEERS (\citealt{bisigello2025helmsdeephighlyextincted}; triangles with gold contours; SED-derived metallicities, hence more uncertain) and CEERS-14821 (\citealt{Bisigello_2025}; black contours; spectroscopic metallicity).}
\label{fig:3scalingrelations}
\end{figure*}

\section{Method}
\label{sec:method}

\subsection{The dustyGadget model}
We base our theoretical analyses on \texttt{dustyGadget}, a hydrodynamical cosmological code developed within the $\Lambda$CDM framework, consistent with \citealt{Planck2016} ($\mathrm{\Omega_{\Lambda} = 0.6911, \Omega_m = 0.3089, \Omega_b = 0.0486, h = 0.6774}$). 
We summarize here the main features of the model, referring the interested reader to \citet{Graziani_2020} for a more detailed description and to \citet{Di_Cesare_2023} for the presentation and analysis of the most recent cosmological simulation suite in the context of high-$z$ galaxy scaling relations, their cold and enriched gas content \citep{carbonenvelopes_DG}. For the characterization and observability of Population III star forming regions in the outskirts of massive galaxies at $z < 10$, we refer to \citealt{Venditti2023, Venditti2026}. 

\texttt{dustyGadget} is built upon the SPH code Gadget-2/3 \citep{Springel2005, Springel2021} and incorporates a self-consistent treatment of dust production and evolution. The code extends the chemical enrichment framework developed by \citet{Tornatore2007a, Tornatore2007b} and the non-equilibrium molecular chemistry and cooling prescriptions introduced by \citet{Maio2007}. In particular, the chemical evolution module follows the release of metals from stars of different masses, metallicities, and lifetimes, as described by \citet[see also \citealp{Maio2010}]{Tornatore2007a}. 
The simulations do not include the formation and evolution of active galactic nuclei, and therefore neglect both mechanical and radiative feedback from accreting supermassive black holes. Photo-heating is implemented through the standard Gadget prescription for the cosmic ultraviolet background originally introduced by \citet{HaardtMadau1996}.

Starting from neutral primordial gas at $z = 100$, the evolution of all particle species is followed down to $z = 4$. The {\sc AMIGA Halo Finder} (AHF; \citealt{AMIGA}) identifies bound structures in post processing. Throughout this work, we use the term \textit{galaxy} to refer to the entire baryonic content enclosed within the virial radius of an AHF-identified halo. \footnote{This is an operational definition, adopted consistently throughout our analyses. At $z \sim 4.5$ structures are still assembling, so the baryonic content within the virial radius may consist of several distinct components, even spatially well separated, especially at lower masses. The \textit{galaxy} here represents  the whole bound system.}

As the simulation evolves, each gas particle tracks the properties of a two-phase interstellar medium, consisting of cold and hot gas phases and their associated dust components. The cold phase has an averaged temperature of $\mathrm{ \overline{T_c}=8000 K} $, while the hot phase spans $\mathrm{T_h \sim 10^4-10^6 K}$ \citep{SpringelandHernquist2003}.

In addition, \texttt{dustyGadget} follows the evolution of several metal species (C, Ca, O, N, Ne, Mg, S, Si, Fe, Na, Al, Ar and Ni) and multiple dust populations, including carbonaceous grains, silicates, alumina, and iron grains. Dust grains are assumed to be spherical and characterized by a fixed representative size of $\approx 0.1 \mathrm{\mu m}$ \citep{Hirashita2014}. 

The dust mass is governed by the combined effects of stellar dust production, grain growth, and destruction processes, including supernova (SN) shocks, sputtering, and astration. These mechanisms are described by the following equation:
\begin{equation}
\mathrm{\dot{M}_{dust}= \dot{M}_{stardust} + \frac{x_c M_{dust}}{\tau_{gg}} - (1- x_c) M_{dust} \left(\frac{1}{\tau_d} + \frac{3}{\tau_{sp}} \right) - SFR(t) D_c}
\end{equation}
where $\mathrm{\dot{M}_{stardust}}$ represents the stellar dust injection rate, $\mathrm{x_c}$ is the cold-gas mass fraction, and $\mathrm{D_c}$ denotes the dust-to-gas ratio in the cold phase. The characteristic timescales $\mathrm{\tau_{gg}}$, $\mathrm{\tau_d}$, and $\mathrm{\tau_{sp}}$ correspond to grain growth, SN-shock destruction, and sputtering, respectively. Stellar dust production is computed according to

\begin{equation}
    \mathrm{M_{stardust, Z_i}(t)= \int_{0}^{t} dt' \int_{m_{\tau_m}}^{m_{up}} m_{dust}(m,Z_i) \phi(m) SFR_{Z_i}(t'- \tau_m) dm}
    \label{eq:stardustmass}
\end{equation}

\noindent where the subscript $Z_i$ denotes a fixed metallicity bin,  $\mathrm{m_{dust}(m,Z_i)}$ is the stellar dust yield as a function of mass and metallicity (here, in the simulation, we adopt \citealt{FerrarottiandGail2006} for AGB stars and \citealt{BianchiandSchneider2007} for supernovae, including reverse-shock processing), $\mathrm{\phi(m)}$ is the initial mass function (IMF) with upper mass limit $m_{up}$, and $\mathrm{SFR_{Z_i}(t'- \tau_m)}$ is the star formation rate evaluated at the formation epoch of stars that end their lives (releasing their dust yield) at time t', having lifetime $\tau_m$ and mass $m_{\tau_m}$.

Grain growth is assumed to occur in unresolved dense structures associated with molecular clouds at $\mathrm{T_c=50 \mathrm{K}}$. The corresponding accretion timescale \footnote{ \label{footnote:gg_timescale} For the definition of the timescales for SN-shock destruction and sputtering, $\mathrm{\tau_d}$ and $\mathrm{\tau_{sp}}$, see \citet{Graziani_2020}. These timescales are typically longer than $\tau_{gg}$ (see, e.g, \citealt{Dwek1998}) and are therefore not discussed in detail here. } is given by
\begin{equation}
\mathrm{\tau_{gg}= 2 Myr \cdot \left( \frac{n_c}{1000 cm^{-3}}\right)^{-1} \cdot \left( \frac{T_c}{50 K}\right)^{-1/2} \cdot \left(\frac{Z_c}{Z_{\odot}} \right)^{-1}}
\label{eq:gg_timescaleDG}
\end{equation}
thereby effectively linking the efficiency of grain growth to the local gas density and metallicity \footnote{The reference grain growth timescale of $2\,\rm Myr$ is derived from the rate of metal accretion onto spherical dust grains with a characteristic size of $0.1\,\rm \mu m$, assuming typical cold neutral medium (CNM) conditions of $n_{\rm c}=10^3\,\rm cm^{-3}$, $T_{\rm c}=50\,\rm K$, and solar metallicity, following the microphysical model of \citet{Asano_2013_grainsize}.}.

The simulation volume and resolution is identical to that adopted in \citet{Di_Cesare_2023}. 
Specifically, we use one of the eight statistically independent cosmological simulations (run U6), which corresponds to a cubic cosmological volume with a side length of $50 \, h^{-1} \,\mathrm{cMpc}$ ($\sim 74,\mathrm{cMpc}$), sampled with $2 \times 672^3$ particles. This implies mass resolutions of $5.2 \times 10^7 \,\mathrm{M_\odot}$ and $8.2 \times 10^6 \,\mathrm{M_\odot}$ for dark matter and gas particles, respectively.

In the adopted simulation runs, star formation is initiated when a gas particle exceeds a density threshold of $132 \mathrm{h^{-2}cm^{-3}}$ ($\mathrm{\sim 287.7 cm^{-3}}$), following \citealt{Maio2010}. At the mass resolution of the present simulation suite, each newly formed stellar particle has an individual mass of $\mathrm{ 1.9 \,\times\, 10^6 M_{\odot}}$ and represents an entire stellar population, inheriting the metallicity of its parent gas particle. Stellar populations are classified as Population (Pop) III when $\mathrm{Z<Z_{threshold} = 10^{-4} \, Z_{\odot}}$ \citep{Maio2007}, for which a Salpeter IMF (\citealt{Salpeter}) in the mass range $[100-500 ]\mathrm{M_{\odot}}$ is assumed \citep{Maio2016}. More metal-enriched populations (Pop II/I) instead follow a Salpeter IMF spanning $[0.1-100 ]\mathrm{M_{\odot}}$.

\subsection{Selected galaxy sample}

Following \citet{Di_Cesare_2023}, we select galaxies at redshift $z \sim 4.5$ with $\log(M_{\star}/\mathrm{M_{\odot}} ) > 8$ and $\mathrm{SFR} > 1\,\mathrm{M_{\odot}}\mathrm{yr}^{-1}$, reaching maximum values of $\log(M_{\star}/\mathrm{M_{\odot}})\sim 10.8$ and $\mathrm{SFR} \approx 200 \,\mathrm{M_{\odot}}\mathrm{yr}^{-1}$ in the resulting sample. 

From this parent sample, we select five galaxies for detailed analyses based on their location within the dust-to-stellar mass relation, aiming to capture key regions of the emerging S-shaped behaviour. Even the least massive of the five selected galaxies contains $274$ stellar particles, which we consider sufficient for the analyses presented here.

Among them, the two most massive systems (G22, G5) were chosen to represent ALPINE-like galaxies, while the other three systems (G818, G3073, and G1910) were selected to probe the origin of the scatter in dust content at fixed stellar mass.
The full population of galaxies at $z \sim 4.5$ is shown as grey points in Fig.\ref{fig:3scalingrelations}, while the selected systems are highlighted with coloured symbols. A summary of their properties is provided in Table \ref{tab:summarytable}.

\begin{figure*}
   \centering

   \includegraphics[width=0.8\textwidth]{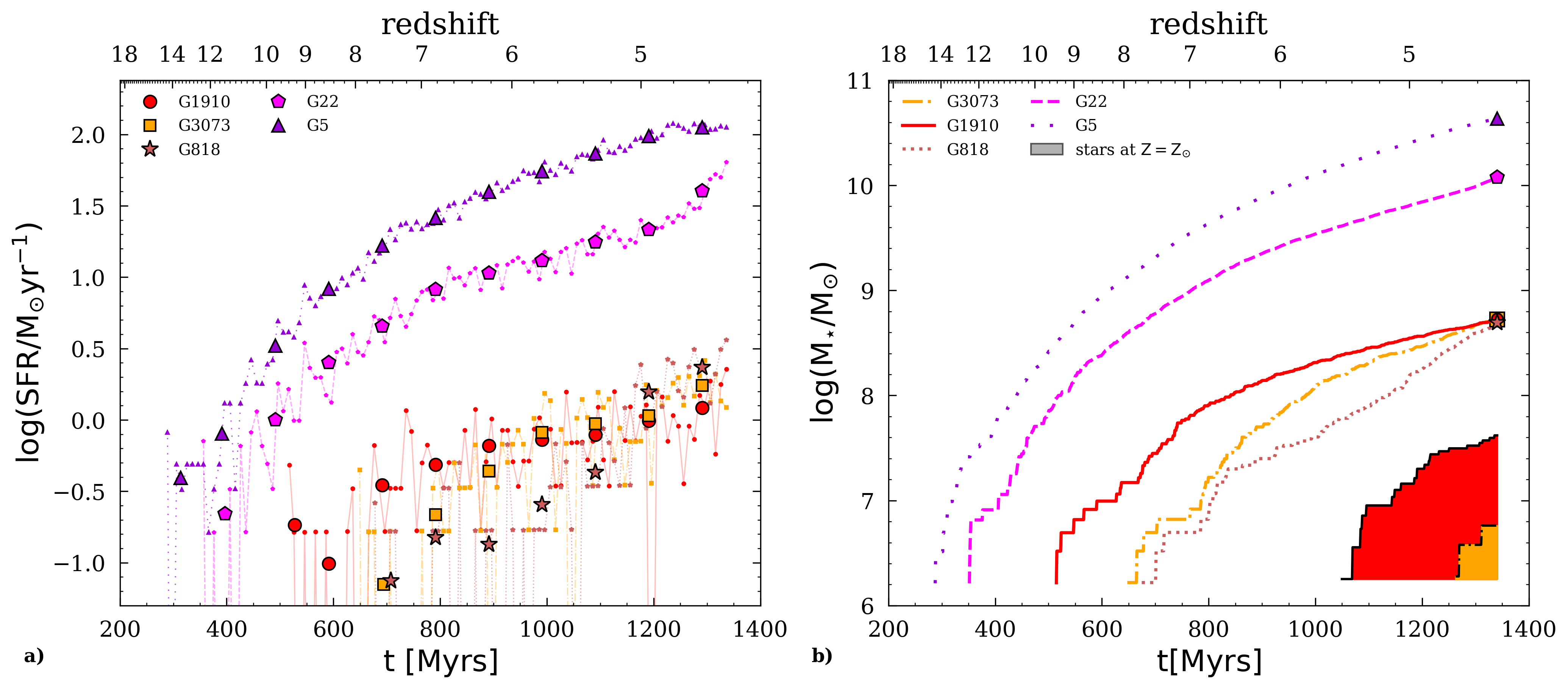}
   \caption{ \textbf{Panel (a)}: Star formation histories averaged over timescales comparable to those adopted in observational studies. The same colour-coding and symbols as in Fig.~\ref{fig:3scalingrelations} are used for each galaxy: coloured dots connected by dotted lines represent $\mathrm{SFH_{10Myr}}$, while the corresponding symbols denote $\mathrm{SFH_{100Myr}}$.
   \textbf{Panel (b)}: Cumulative stellar mass assembled over time, reconstructed from $\mathrm{SFH_{1Myr}}$. The total stellar mass predicted by dustyGadget is shown for the five galaxies using the same colour-coding as in Fig.~\ref{fig:3scalingrelations}, with different line styles distinguishing individual systems. For G3073 and G1910, the contribution from the most metal-rich stellar populations ($Z = Z_{\odot}$) is shown as black curves with the corresponding line styles, while their cumulative contribution is highlighted by shaded regions following the colour assigned to each galaxy. No stars with solar metallicity are present in G818.}
    \label{fig:SFH_complete}           
\end{figure*}

\section{Results}
\label{sec:results}

\subsection{Selection within key scaling relations}
\label{subsec:scalingrelations}

We begin by summarizing the key scaling relations relevant to our study of dust abundance in early galaxies (Fig.\ref{fig:3scalingrelations}). 
While our analysis is anchored in the dust-to-stellar mass relation (panel a), we also present the dustiness-metallicity relation (panel b\footnote{Following \citealt{Galliano_anearbyperspectiveondustevolution}, we refer to the dust-to-gas mass ratio as \textit{dustiness}.}) and the mass-metallicity relation (MZR; panel c). Together, these relations provide a broader interpretative framework, enabling more robust comparisons with both existing studies and forthcoming observations.

In the dust-to-stellar mass plane, the selected galaxies (G818, G3073, and G1910) span nearly the entire range of dust-mass scatter within the central region of the diagram. The remaining systems (G22 and G5) occupy the high-stellar-mass regime, where observational constraints from the ALPINE sample are available (\citealt{Pozzi_sshape}, \citealt{Sommovigo_sshape}) and are broadly consistent with our predictions.

For the dustiness--metallicity relation, we compare our results with the best-fit relation derived for nearby galaxies by \citet{Galliano_anearbyperspectiveondustevolution}. 
We use the mean gas-phase metallicity estimated from \texttt{dustyGadget}: for each young star particle ($<10 \rm Myr$), the code estimates the metallicity of the gas within a surrounding sphere approximating the expected \textsc{H\,ii} region, and the galaxy is then  assigned the mean value across all such regions, making our estimate as comparable as possible to observational determinations.

Following \citet{Asanoetal2013}, we also highlight the metallicity regimes associated with the dominant dust production mechanisms: the stardust regime at $\mathrm{12 + \log(O/H)} < 7.5$, the grain growth regime at $\mathrm{12 + \log(O/H)} > 8.2$, and an intermediate transition region, where both channels contribute owing to differences in galaxy star formation histories.

The simulated galaxies at $z \sim 4.5$ exhibit a similar overall behaviour, although the characteristic S-shaped trend is less pronounced than that observed in the local Universe. The mean relation (black dotted line), together with its dispersion (violet shaded region), suggests a change in slope around $\mathrm{12 + \log(O/H)} \sim 7.3$ ($\rm Z\sim \, 0.04 Z_{\odot}$, based on the solar abundance from \citep{Asplund2009}), slightly below the transition metallicity predicted by \citet{Asanoetal2013}\footnote{We note, however, that the precise value of this transition depends both on the adopted dust-evolution model and on the definition of a \textit{galaxy}. Unlike the well-defined galaxy structures typically considered at $z=0$, galaxy identification at high redshift is less straightforward and in our simulations includes contributions from the circumgalactic medium (CGM).}.

The locations of the five selected galaxies in this plane closely reflect their dust content, with dust-poor systems systematically occupying lower dustiness values. G818 is the least metal-enriched galaxy and lies close to the metallicity at which the average relation of the simulated sample changes slope. Interestingly, G3073 has a metallicity only $\simeq 0.1$ dex higher than G818, yet its dustiness is larger by nearly 1 dex. By contrast, the difference in dustiness between G3073 and G1910 ($\simeq 0.4$ dex) closely matches their metallicity difference. This suggests that, at these low stellar masses, metallicity alone is not a reliable predictor of dustiness. Finally, G22 and G5 lie close to the relation defined by local galaxies of comparable metallicity.

A similar picture emerges from the mass--metallicity relation, where G818/G3073, and G1910 exhibit a significant difference in metallicity despite having comparable stellar masses.

  Overall, the simulated galaxies reproduce the observed mass-metallicity relation across the full stellar mass range, showing particularly good agreement at the massive end \citep{Faisst_massmetallicity}. At lower stellar masses, however, the simulated systems tend to be more metal-poor than the currently available observational samples (\citealt{Curti_massmetallicity}; \citealt{bisigello2025helmsdeephighlyextincted}; \citealt{Bisigello_2025}; \citealt{Nakajima_massmetallicity}; \citealt{Chakraborty_massmetallicity}).

  In panel (c) we compare the mass-metallicity relation from our simulations to that from observations at z = 3 - 6. These rely on different calibrations to estimate the gas-phase metallicity (e.g. strong line calibration or the direct temperature method; see \citealt{Maiolino&Mannucci2019,Curti2026review}) and on different selection functions (see \citealt{Kotiwale2026} for a discussion of the importance of selection effects). Observations of the MZR reveal substantial scatter across the stellar-mass range, and there is still no consensus on its slope at high redshift \citep{Kotiwale2026, Nakajima_massmetallicity, Curti_massmetallicity, Stanton2026}. Our simulated galaxies show particularly good agreement with observations at the massive end \citep{Faisst_massmetallicity}. At lower stellar masses, however, the simulated systems tend to be more metal-poor than those in currently available observational samples \citep{Curti_massmetallicity, bisigello2025helmsdeephighlyextincted, Bisigello_2025, Nakajima_massmetallicity, Chakraborty_massmetallicity}.

In this context, we highlight CEERS-14821, a spectroscopically confirmed dusty dwarf galaxy at $z \sim 5$ \citep{Bisigello_2025}. Under different dust attenuation prescriptions (\citealt{C00} and \citealt{CF00}), the R3 metallicity diagnostic\footnote{The R3 metallicity diagnostic, defined as $\mathrm{R3}=  \mathrm{log(\frac{[OIII] \lambda 5007}{H\beta}})$, provides a well-behaved relation with gas-phase metallicity while being uneffected by dust reddening, although it strongly depends on the ionization parameter \citep{Maiolino&Mannucci2019}.} yields two self-consistent solutions, corresponding to lower- and higher-stellar-mass estimates with high- and low-metallicity values, respectively. Given its redshift and stellar mass, particularly under the \citet{CF00} attenuation model, CEERS-14821 represents a plausible observational analogue of G818 and G3073.
 
Motivated by the correspondence between the dustiness trends observed in local galaxies and the scaling relations followed by our simulated galaxies, we associate the three dust formation regimes with distinct stellar mass ranges, approximately defined by the transitions seen in these relations (arrows in Fig.~\ref{fig:3scalingrelations}a). This association is further supported by the quantitative analysis presented below.
In the following, we assess the relative efficiency of the two dust production mechanisms and investigate how their interplay regulates the transitions along the scaling relations.

\subsection{Star Formation Histories}
\label{subsec:SFHs}

We reconstruct the individual star formation histories (SFHs) of the selected galaxies by binning their stellar particles according to their formation times. This retrospective approach, starting from the final simulation snapshot at $z = 4.5$ ($t \sim 1.34\,\mathrm{Gyr}$), allows us to reconstruct the recent star formation activity with a flexible temporal resolution. Indeed, since stellar particles are created at the hydrodynamical timesteps of the simulation, their formation times can be binned a posteriori using a suitably chosen time resolution (see \citealt{Venditti2023}).

Figure \ref{fig:SFH_complete} presents the SFHs computed using time bins of  $10 \,\mathrm{Myr}$ ($\mathrm{SFH_{10Myr}}$) and $\mathrm{100 \,Myr}$ ($\mathrm{SFH_{100Myr}}$, panel a), together with the cumulative stellar mass growth at $\mathrm{1 \, Myr}$ resolution ($\mathrm{SFH_{1Myr}}$, panel b).
The $100 \, \mathrm{Myr}$ binning is best suited for comparison with observationally inferred star formation rates, and the corresponding values are reported in Table \ref{tab:summarytable}.
Throughout the following analysis, however, we adopt the finer $1\,\mathrm{Myr}$ resolution, which is more consistent with the intrinsic burstiness of PopIII formation events, as Population III stars have mass averaged lifetimes are of order $\sim 3 \,\mathrm{Myr}$ \citep[obtained by integrating the stellar lifetimes of \citealt{Schaerer2002} over our top-heavy IMF; see][]{Venditti2023}.

The selected galaxies exhibit a wide diversity of star formation histories, differing in their onset, duration, and overall shape. In particular, the two most massive and dust-rich systems (G22 and G5) follow remarkably similar evolutionary paths, whereas G818, G3073, and G1910 display markedly different SFHs despite converging to nearly identical stellar masses by the final snapshot.

For the latter three systems, we further exploit the stellar metallicity to reconstruct SFHs restricted to a near-solar metallicity bin (solid black curves with color-coded shading underneath for each galaxy). Because stars inherit the metallicity of their parent gas at birth, these histories provide a direct probe of the chemical enrichment required for efficient grain growth in star-forming regions. They reveal that, despite their similar final stellar masses, the three lower-mass galaxies begin to diverge in gas-phase metallicity already at $z \lesssim 6$, consistent with the metallicity spread shown in Fig.~\ref{fig:3scalingrelations}c.

\begin{figure*}
   \centering
   
   \includegraphics[width=0.8\textwidth]{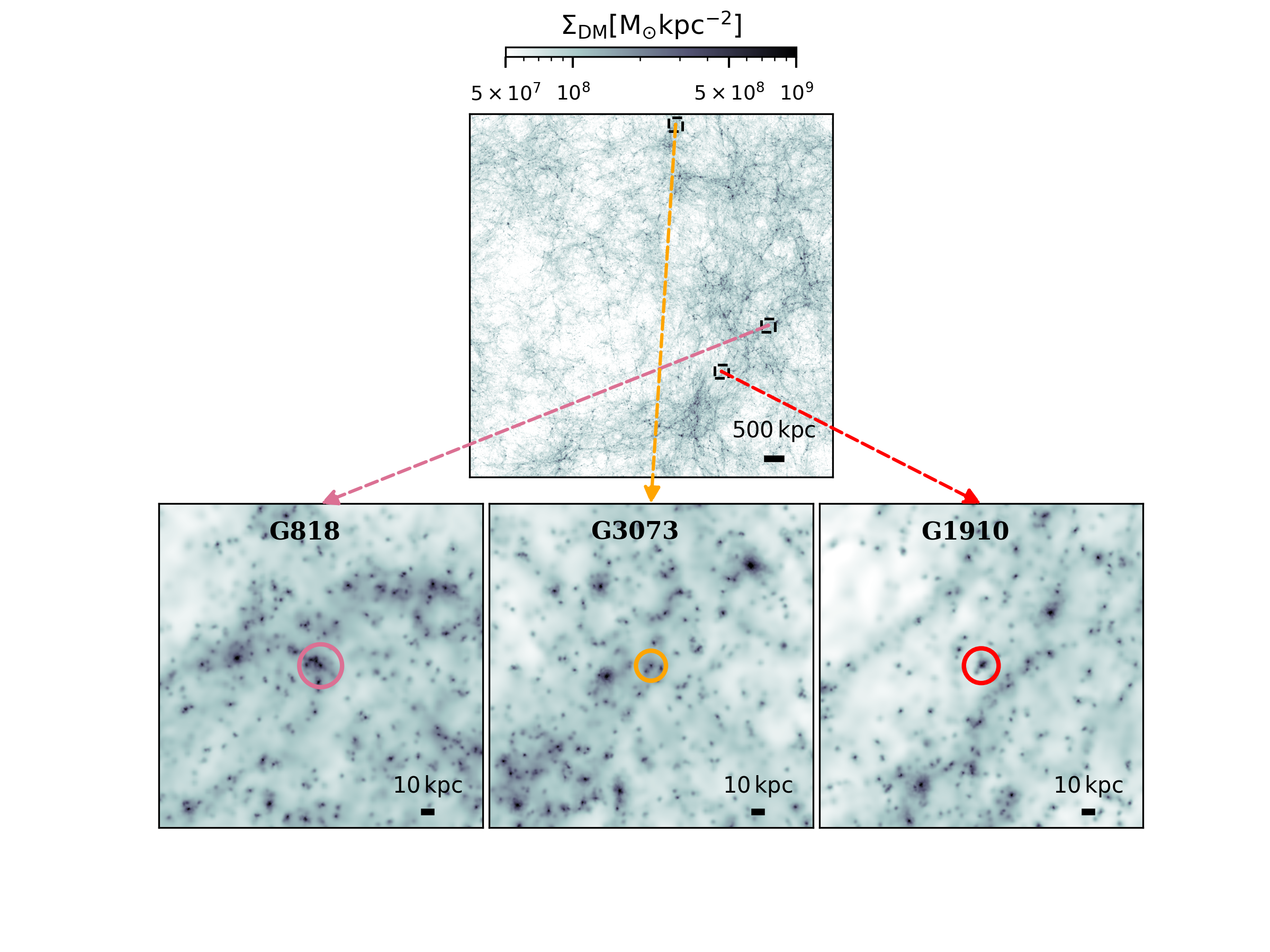}
   \caption{ \textbf{Upper panel}: dark matter surface density map computed by projecting the full cosmological volume. 
   The three transition-region galaxies are highlighted with circles indicating their virial radii. \textbf{Lower panel}: zoom-in of the upper panel, showing the local environments of the three transition-region galaxies.}
    \label{fig:cw}              
\end{figure*}

\begin{figure*}
\centering

\includegraphics[width=0.5 \textwidth]{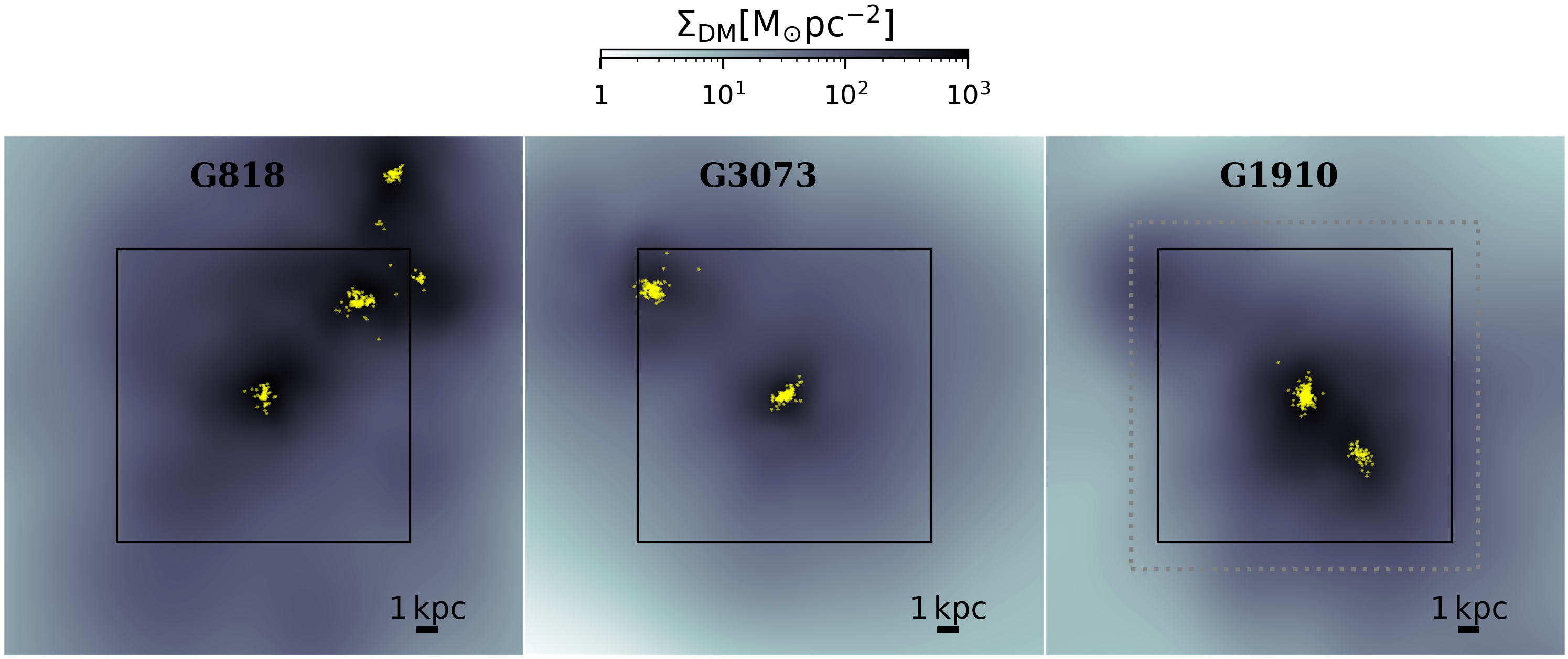}\\[-1mm]
\includegraphics[width=0.5\textwidth]{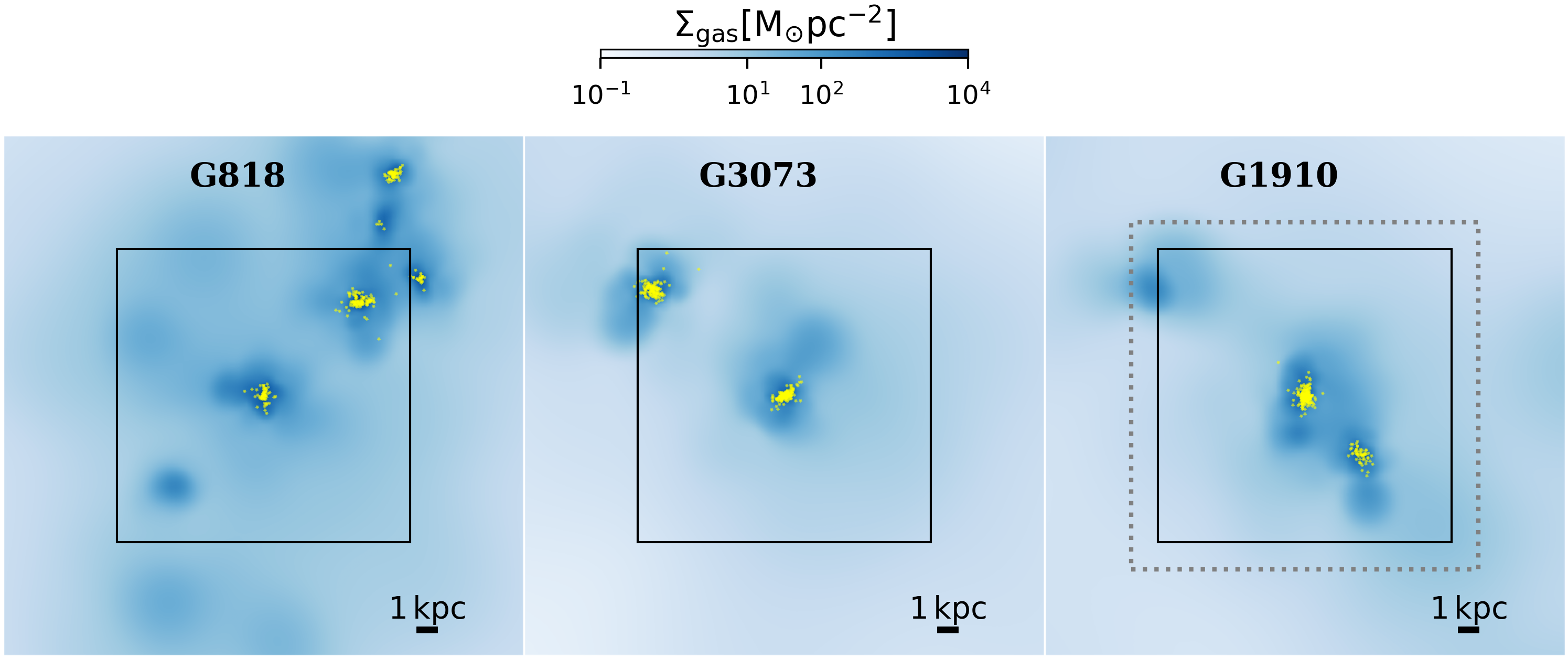}\\[-1mm]
\includegraphics[width=0.5 \textwidth]{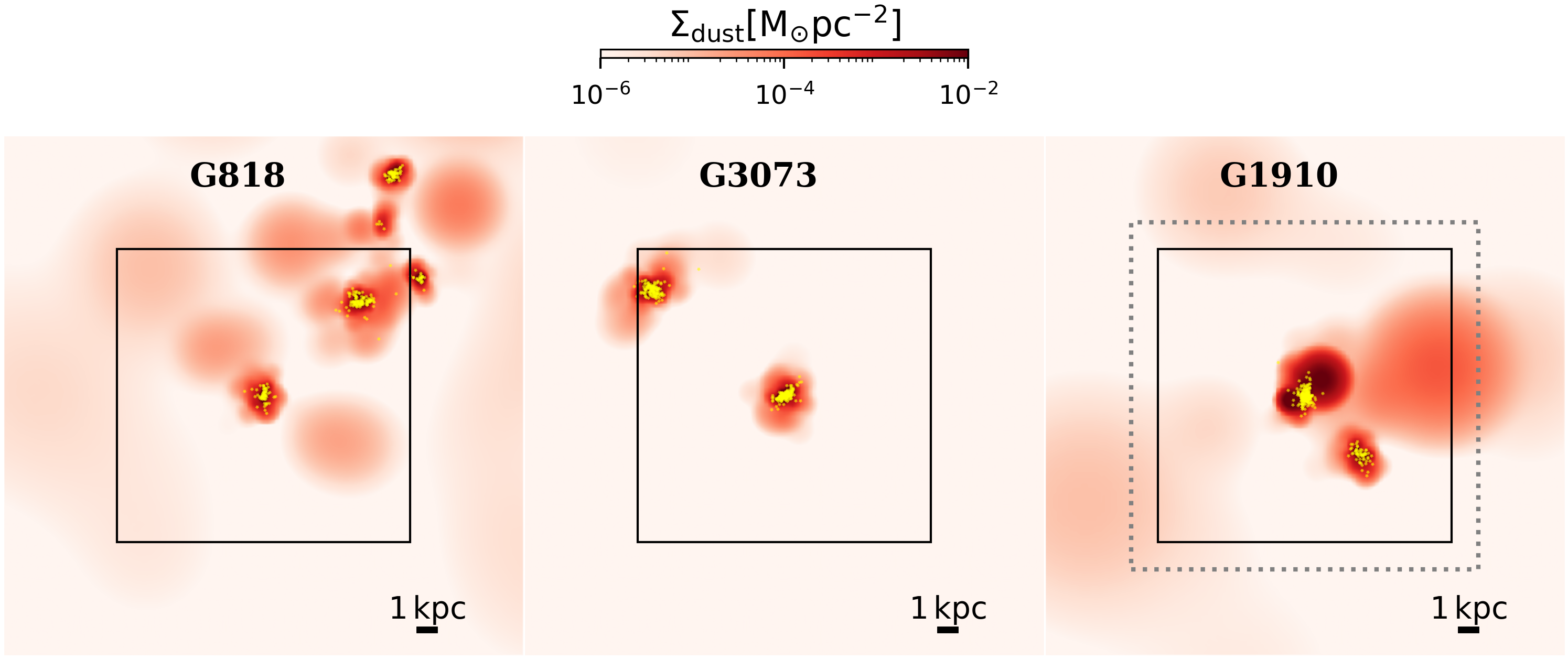}

\caption{The top, middle, and bottom panels show the dark matter, gas, and dust surface density maps, respectively. Each column corresponds to a 
different galaxy and spans the same physical size, chosen to encompass its entire stellar structure. Stellar particles are overplotted as 
yellow dots. For direct comparison with our observational predictions (Figs.~\ref{fig:JWSTmock} and \ref{fig:ALMAmock}), the black square 
marks the common $\pm 1.5\,\mathrm{arcsec}$ region shown in the JWST mock zooms, encompassing the two main clumps of each galaxy. The wider 
$\pm 3\,\mathrm{arcsec}$ field adopted for the full JWST and ALMA mocks extends beyond the area displayed in these panels. For G1910, the grey 
dotted square additionally marks the extent of the SKIRT input cube; for the other galaxies, the corresponding region is too large to be 
shown at the scale of this figure. See Figs.~\ref{fig:spectra_focusonspectralanalyses} and \ref{fig:AVcomparisons} for further analyses based on 
the SKIRT radiative-transfer calculations.}
\label{fig:grids}
\end{figure*}

\subsection{Stardust contribution}

To isolate the stellar contribution to the total dust mass predicted by \texttt{dustyGadget}, we employ a semi-analytical chemical evolution model. This enables us to assess whether the physical interpretation of the slope transition in the dust-to-stellar mass relation, originally established for local galaxies, remains valid at high redshift. The time evolution of the stellar dust mass is computed following Eq. \ref{eq:stardustmass}, adopting the same Salpeter initial mass function and dust yields as those implemented in the simulation (see Section \ref{sec:method})), with stellar lifetimes estimated following the parametric form proposed by \citealt{Raiteri1996}.

The model follows the evolution of stellar dust mass in five metallicity bins, spanning $10^{-4}\,Z_{\odot}$ to $1\,Z_{\odot}$, using the star formation histories derived from \texttt{ dustyGadget} as input. The total stellar dust mass is then obtained by summing the contributions from stellar populations formed at each metallicity.

Table~\ref{tab:summarytable} summarizes, for each galaxy, the total dust mass predicted by \texttt{ dustyGadget}, together with the stellar dust contribution estimated by our semi-analytical model ($M_{\rm stardust}$) and the fraction of stellar dust produced by AGB stars alone ($f_{\rm AGB}$).

Focusing on the S-shaped trend in the dust-to-stellar mass relation (Fig.~\ref{fig:3scalingrelations}a), we fit the lowest part of the low-mass branch (grey dash-dotted line), assuming that dust production in this regime is dominated by 
stellar sources, while dust destruction and astration provide only a minor contribution to the dust budget. This assumption is motivated by 
the fact that both dust removal rates are directly proportional to the dust-to-gas ratio \citep[e.g.,][]{sorgentidipolvere}. Indeed, at fixed 
low metallicity, the galaxies included in the fit exhibit the lowest dust-to-gas ratios in our sample, naturally suppressing both destruction 
and astration and allowing the stellar dust contribution to dominate. This interpretation is further supported by the close agreement between 
the low-mass fit and the stellar-only predictions (empty symbols). Together, these results guide our identification of the three dust 
formation regimes. In particular, we define the transition regime as the region where the simulated relation begins to deviate from the 
extrapolation of the low-mass fit. Within this region, G818, G3073, and G1910 are predicted to have comparable stellar dust contributions.

At the final redshift considered, $z=4.5$, the contribution of AGB stars to $M_{\rm stardust}$ is significant, although it remains sub-dominant, 
ranging from $f_{\rm AGB}=0.32$ for G818 to a maximum of $f_{\rm AGB}=0.46$ for G1910. Interestingly, this highest contribution 
is not found in the most massive galaxy (G5), indicating that the relative roles of SNe and AGB stars in stellar dust production are 
governed more by the galaxy's SFH than by its cumulative stellar mass growth.

The relative contribution of AGB stars is known to depend sensitively on the adopted IMF, the onset of star formation (which 
sets the time available for stars to evolve into the AGB phase) and the adopted metallicity-dependent stellar yields \citep{sorgentidipolvere}. 
For the two most massive galaxies (G5 and G22), which begin forming stars at $z \sim 14$, we find that AGB stars 
contribute $<20\%$ of the stellar dust mass at $z>7.5$ and $<10\%$ at $z>10$, with even smaller fractions found for the three less massive 
galaxies. We stress that these fractions refer to the stellar dust  component, $M_{\rm stardust}$, which is itself sub-dominant with respect 
to the total dust mass. The contribution of AGB stars to the total dust budget is therefore correspondingly smaller.

\subsection{Environment and morphology}

From here on, we focus on the three galaxies located in the transition region, whose differing dust content is responsible for the scatter seen in Fig.\ref{fig:3scalingrelations} (panel a) \footnote{  Whenever the other galaxies are discussed in the following, their properties are summarised in Tables, and the galaxy-specific figures are provided in the Appendix. }. To further investigate the origin of these differences, we examine their location within the cosmic web and analyse their internal morphology. 

Figure \ref{fig:cw} first presents the full cosmological volume as a dark matter surface density map (upper panel), with the three transition-region galaxies marked by circles corresponding to their virial radii\footnote{The maps shown in Figs. \ref{fig:cw} and \ref{fig:grids} are generated using the {\sc pynbody} framework \citep{pynbody_Pontzen2013}, which is employed to process the simulation outputs and produce projected mass surface density maps.}. The lower panel shows zoomed-in views of their local environments. All three systems reside in dense regions associated with cosmic web filaments. Moreover, we find no significant differences in their large-scale environments, as the matter density on scales of $\sim500\,\mathrm{kpc}$ -$1\,\mathrm{Mpc}$ is comparable across the sample, with mean values of $\sim 1.7$ and $\sim 1.3$ times the cosmic mean density, respectively.

We next examine the internal structure of these galaxies in Fig.~\ref{fig:grids}, which compares the projected surface density distributions of dark matter, gas, and dust, with stellar particles overlaid in each panel. All systems correspond to the baryonic component associated with a single dark matter halo identified by the {\sc AMIGA Halo Finder}. The dark matter distribution (top panels) is highly structured, with stellar substructures tracing the underlying subhaloes. In contrast, the gas (middle panels) and dust (bottom panels) exhibit more extended and diffuse morphologies, likely reflecting the effect of feedback processes, although the smoothing scales adopted to build the maps may also contribute. While dust remains spatially correlated with the stellar substructures, it extends well beyond them, often reaching regions with no obvious stellar counterpart.

The surface density maps also reveal significant differences in the degree of substructure among the three systems. To quantify these morphological differences, we identify the individual stellar clumps and measure their contributions to the total stellar and dust mass. G818 exhibits the most fragmented configuration, with several substructures. These clumps contribute between $\sim 2\%$ and $\sim 40\%$ of the total stellar mass, and from $<0.1\%$ up to $\sim 35\%$ of the total dust mass. G3073 is dominated by two major components, each contributing nearly equally to the stellar ($\sim50\%$) and dust ($\sim40\%$) mass. In both G818 and G3073, the substructures extend to projected distances exceeding $10 \rm kpc$ from the halo centre, highlighting the spatially extended nature of their baryonic distributions. In contrast, G1910 appears to represent a more advanced merger remnant, with two components; the secondary clump contributes $\sim20\%$ of the stellar mass but less than $1\%$ of the total dust budget. 

We therefore conclude that, at these early stages, particularly in the smaller systems, the baryonic content within a single dark matter may still be distributed among multiple internal clumps, reflecting different stages of the assembly process. Each baryonic substructure may undergo local gravitational collapse and host its own star formation activity. On these small scales, the stochastic nature of the assembly process implies that the system has not yet developed a single, coherent stellar component or a well-structured, multiphase ISM with dense gas concentrations. Such features, which are more typical of evolved systems, provide the physical conditions conducive to efficient star formation and grain growth.

\subsection{Synthetic spectra}

\begin{figure*}
   \centering
   \includegraphics[width=0.7\textwidth]{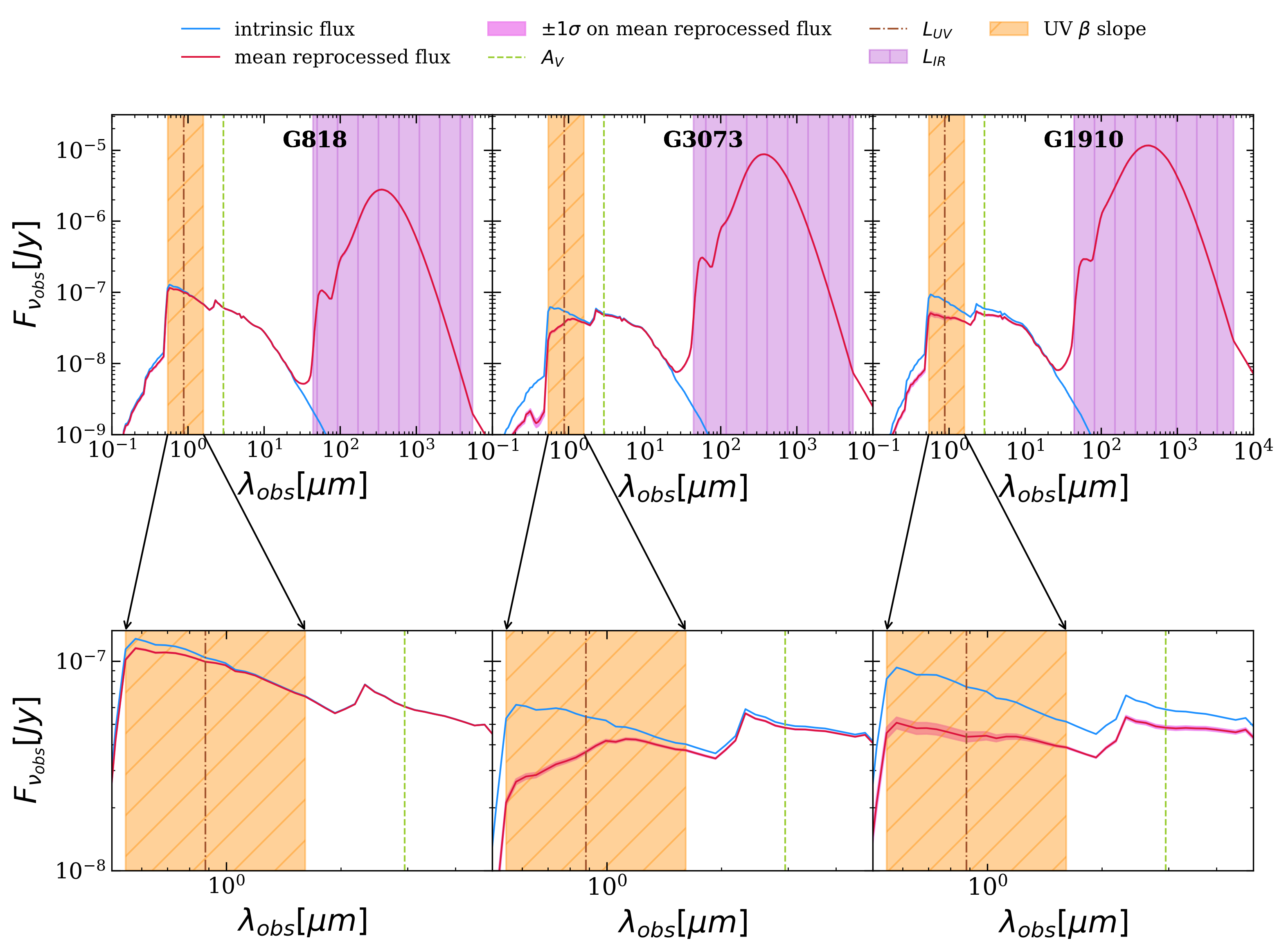}
   
   \caption{Synthetic SEDs of the selected galaxies at $z\sim4.5$, located in the transition region of the dust-to-stellar mass relation. The bottom panel provides a zoom-in of the wavelength range dominated by stellar emission, highlighting the spectral region most relevant for attenuation diagnostics. The orange shaded region marks the wavelength interval used to estimate the UV continuum slope, $\beta$, while the green dotted line indicates the central wavelength of the rest-frame $V$ band, where the attenuation, $A_V$, is derived by comparing the intrinsic and dust-reprocessed fluxes. The intrinsic stellar emission is shown by the blue solid curve, whereas the mean dust-reprocessed SED, averaged over different lines of sight, is shown by the red solid curve. The violet shaded region indicates the corresponding $\pm1\sigma$ dispersion. The brown dash-dotted line marks the observed-frame wavelength corresponding to rest-frame $0.16\,\mu$m, adopted for the derivation of $L_{\rm UV}$, while the purple shaded region indicates the wavelength range used to estimate $L_{\rm IR}$.
   }
              \label{fig:spectra_focusonspectralanalyses}%
              
    \end{figure*}

\begin{table*}

\centering
\caption{Dust-related diagnostics derived from the synthetic spectra of the selected galaxies, including systems spanning the dust transition region (G818, G3073, and G1910) and the more massive galaxies (G22 and G5). All quantities, except for the intrinsic UV continuum slope, $\beta_{\rm int}$, are derived from the dust-reprocessed spectra averaged over multiple lines of sight and are reported as mean values together with their associated uncertainties. See the caption of Fig.~\ref{fig:spectra_focusonspectralanalyses} for the definition of the reference wavelength ranges adopted to derive the reported quantities.}

\resizebox{\textwidth}{!}{

\begin{tabular}{ |c||c|c|c|c|c||c|c|c|c|c|}

 \hline
 
 &$\mathrm{\beta_{int}}$ 
 &$\mathrm{\beta_{rep}}$ 
 &$\mathrm{L_{IR}[L_{\odot}]}$
 &$\mathrm{L_{UV}[L_{\odot}]}$
 &$\mathrm{M_{UV} [mag]}$
 &$\mathrm{IRX}$
 &$\mathrm{A_V}$\\
 
 \hline
 \textbf{G818} &-2.576 & $-2.467 \pm 0.001$ & $(2.122 \pm 0.003) \cdot 10^{9}$ & $(1.872 \pm 0.001)\cdot 10^{10}$& $-21.716 \pm 0.001$ &$ (1.137 \pm 0.002) \cdot 10^{-1}$& $(2.5 \pm 0.2) \cdot 10^{-3} $\\
 \textbf{G3073} &-2.391&$-1.53 \pm 0.01$  & $(6.32 \pm 0.01) \cdot 10^{9}$ & $(6.94\pm 0.03) \cdot 10^{9}$& $-20.638 \pm 0.005 $&$ (9.1 \pm 0.1) \cdot 10^{-1}$& $(4.1 \pm 0.1) \cdot 10^{-2}$ \\
 \textbf{G1910} & -2.545 &$ -2.19 \pm 0.02 $ &$(8.919 \pm 0.004) \cdot 10^{9}$& $(8.2 \pm 0.1) \cdot 10^{9}$& $-20.82\pm 0.02$ &$1.09 \pm 0.06$ & $(2.18 \pm 0.05) \cdot 10^{-1}$\\
 \textbf{G22} &  -2.543& $-1.36 \pm 0.04$ & $(5.10 \pm 0.01) \cdot 10^{11}$& $(3.0 \pm 0.1)\cdot 10^{10}$& $-22.22 \pm 0.04$ &$ (1.7 \pm 0.1) \cdot 10^{1}$& $1.57 \pm 0.02$\\
 
 \textbf{G5} & -2.293 & $-1.48 \pm 0.02$ & $(7.71 \pm 0.08) \cdot 10^{11}$ & $(3.12 \pm 0.04)\cdot 10^{10}$& $-22.27 \pm 0.01$&$(2.5 \pm 0.1) \cdot 10^{1}$& $2.47 \pm 0.02$ \\
 \hline
 
\end{tabular}
}
\label{tab:spectralanalyses}
\end{table*}

\begin{figure*}
   \centering
   
   \includegraphics[width=0.8\textwidth]{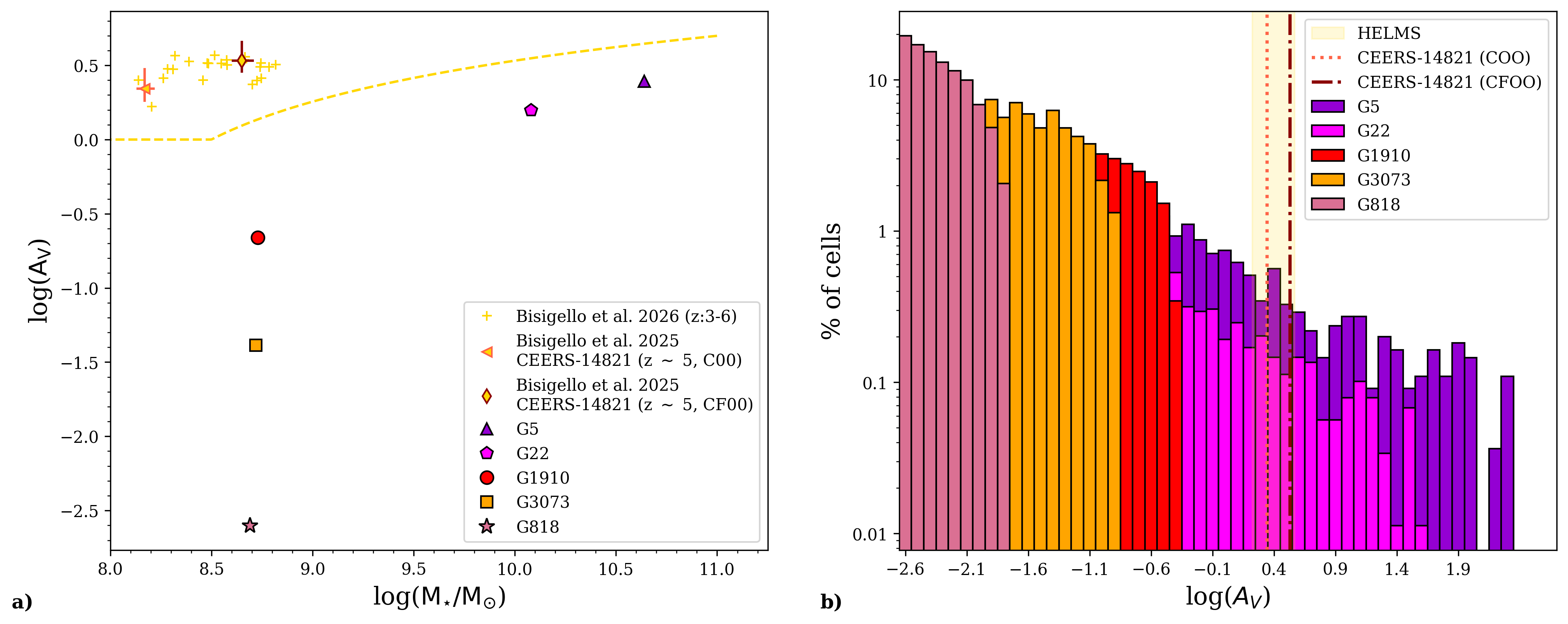}
   
   \caption{\textbf{(a)} Relation between the dust attenuation and the stellar mass for the selected galaxies. Theoretical predictions derived from the synthetic spectra are shown as coloured symbols. The dashed gold line marks the criterion adopted by \citet{bisigello2025helmsdeephighlyextincted} to define Highly Extincted Low-Mass (HELM) galaxies. White triangles with gold outlines denote the observed HELM galaxies at $z\sim 3-6$ from the same work, while the two estimates for CEERS-14821 are shown as gold triangles with black outlines. \textbf{(b)} Spatial distribution of the visual attenuation within the selected galaxies, showing the cell-by-cell $A_V$ values (see text for details on the spatial scales) using the same colour scale as in the main panel. The range of $A_V$ values measured for HELM galaxies at comparable redshifts is highlighted by the yellow shaded region, while the two estimates for CEERS-14821, obtained assuming different dust attenuation models, are shown as black lines.}  
   \label{fig:AVcomparisons}           
\end{figure*}

We reconstruct synthetic spectral energy distributions (SEDs) using a customised version of the radiative transfer code 
{\sc SKIRT} \citep{SKIRT}, modified to directly ingest the dust distribution predicted by \texttt{dustyGadget} for each galaxy, rather 
than inferring it from the gas metallicity. This approach preserves both the total dust mass and its spatial distribution on the computational 
grid, ensuring that the resulting SEDs, as well as the derived UV slopes $\beta$ and $A_V$ values, consistently reflect the dust column densities 
predicted by \texttt{dustyGadget}.

Our setup adopts standard assumptions for stellar population synthesis and dust properties. Specifically, we use the \citet{Bruzual_Charlot_2003} stellar population synthesis models together with the \citet{Draine_2003} dust model, assuming a Small Magellanic Cloud (SMC) extinction curve, which is commonly adopted as a benchmark for high-redshift galaxies. Dust emissivity is computed following the prescription of \citet{Mathis_1977}, accounting for the grain size distribution and composition. 

We neglect nebular emission and model only the continuum component. At our resolution, the birth clouds surrounding young stars ($<10\,\mathrm{Myr}$) are not resolved, and we therefore do not include sub-grid attenuation within these regions. Previous studies that explicitly incorporate sub-grid birth-cloud models and compare radiative-transfer predictions with and without them provide an indication of the potential impact of this assumption, although they do not allow us to quantify it for our specific galaxy sample. In particular, neglecting birth clouds has been found to shift the UV slope $\beta$ towards bluer values 
($\Delta\beta \sim -0.3$) and to reduce the IRX by $\approx 0.1$--$0.2$ on average at $z=0.5$ \citep{Schulz2020}, with the largest effects found in lower-mass systems. It may also lead to a slight underestimate of the MIR emission, while having a negligible effect ($\lesssim 5\%$) on the mass-weighted dust temperature 
\citep{Liang2019}, and has been found to increase the dust-continuum radius by $\approx 10\%$ \citep{Popping2022}. We therefore acknowledge the lack of a sub-grid treatment of birth clouds as a limitation of our modelling. However, since stars younger than $10\,\mathrm{Myr}$ account for only $\sim 2$-$7\%$ of the stellar mass in our sample, we expect the overall impact of unresolved birth-cloud attenuation on the integrated continuum emission and dust energy balance to be modest.

For each galaxy, we extract a galaxy-centred cubic region encompassing the entire stellar distribution and extending slightly beyond it to include the surrounding dust. This choice provides sufficient spatial coverage to capture dust reprocessing while preserving an adequate spatial resolution. The side length of the extracted regions is $\sim 50, 60, 24, 140$ and $160$ physical kpc for G818, G3073, G1910, G22 and G5, respectively, with corresponding dust-grid cell sizes of $\sim 0.2, 0.2, 0.1, 0.5$ and $0.6$ physical kpc. 

We then compute the SEDs along multiple lines of sight for each galaxy, including representative extreme viewing configurations.

Figure~\ref{fig:spectra_focusonspectralanalyses} presents the resulting spectra for the three transition-region galaxies.
To highlight the impact of dust, we compare the reprocessed SEDs with the intrinsic stellar emission (red and blue curves, respectively).
Finally, to assess whether the dust content of these systems can be robustly constrained observationally, we derive a set of diagnostics from the reprocessed emission. For completeness, the corresponding SED-derived quantities for the full galaxy sample are reported in Table~\ref{tab:spectralanalyses}.

In the UV regime, where dust absorption is particularly efficient, we focus on the reprocessed UV luminosity and the dust-induced reddening by comparing UV slopes measured from the intrinsic and reprocessed emission. UV slopes are obtained by fitting the rest-frame flux $F(\lambda) \propto \lambda^{\beta}$, in units of $ \mathrm{erg \cdot s^{-1} \cdot cm^{-2}\cdot A^{-1}} $,  over $[0.1 - 0.3 ]\mathrm{\mu m}$ (orange shaded region).
The intrinsic UV slopes span a relatively narrow range, from $\beta_{\rm int}=-2.576$ for G818 to $\beta_{\rm int}=-2.293$ for G5, indicating that the unattenuated UV continua are comparably blue across the sample, with only modest variations associated with differences in their recent star formation histories and stellar populations. The reprocessed slopes span a much wider range than their intrinsic counterparts, revealing marked differences in dust reddening. In particular, despite its lower dust content, G3073 has a substantially redder UV continuum than G1910, with $\beta_{\rm rep}=-1.53$ compared with $-2.19$, while the strongest reddening is found in one of the more massive systems, G22, for which $\beta_{int}=-2.543$ and it increases by $\Delta\beta\simeq1.18$.
The comparison between G3073 and G1910 demonstrates that the effect of dust attenuation is not determined solely by the total dust mass, but also by the relative distribution of dust and stars. This highlights the importance of galaxy morphology and internal structure when interpreting observed UV colours and attenuation diagnostics \citep{Salim2020, Zhang2023}.

The UV luminosity is estimated from the rest-frame flux at $\lambda \sim 0.16\,\mathrm{\mu m}$ (brown dash-dotted line). Among the transition-region galaxies, G818 is the brightest UV source, consistent with its higher star formation rate and lower dust content (Table~\ref{tab:summarytable}). G3073 and G1910 are progressively fainter, reflecting the combined effect of decreasing SFR and increasing dust content. 

As additional tracers of the dust effect, we compute the visual attenuation, $A_V$, the infrared luminosity, $L_{\rm IR}$, and the infrared excess, $\mathrm{IRX}=L_{\rm IR}/L_{\rm UV}$. The former quantifies the attenuation in the rest-frame V band ($\lambda \sim 0.55\,\mathrm{\mu m}$; green dashed line), while $L_{\rm IR}$ corresponds to the bolometric dust luminosity integrated over the $[8 - 1000]\,\mathrm{\mu m}$ wavelength range. As expected, all three quantities increase systematically with dust content: galaxies with larger dust masses exhibit higher visual attenuation, larger infrared luminosities, and higher IRX values, indicating that an increasing fraction of the stellar radiation is absorbed by dust and re-emitted in the infrared. This trend is reflected in the progressively larger separation between the intrinsic and dust-reprocessed SEDs, the emergence of a more prominent infrared bump, and the increasing line-of-sight scatter at short wavelengths, where dust absorption is most effective.

In Fig.~\ref{fig:AVcomparisons}, we compare the galaxy-integrated attenuation derived from the SKIRT synthetic spectra (coloured symbols) 
with the observational constraints from CEERS-14821 (nebular attenuation, derived from the Balmer decrement) and 
the HELM sample (continuum attenuation, derived from broad-band photometry), see also Fig.~\ref{fig:3scalingrelations}. At face value, 
all simulated galaxies lie below the lower envelope of the attenuation threshold adopted to define HELM galaxies (dashed gold line).

We note that, since we do not model nebular emission, our predictions are methodologically more directly comparable to the 
continuum-based HELM measurements than to the nebular attenuation inferred for CEERS-14821. 
A second, independent difference concerns the physical scales probed by the observations and simulations. 
 
As an illustrative comparison of the spatial scales, we consider CEERS-14821, which is comparable in both stellar mass and redshift to our simulated galaxies G818, G3073, and G1910, and for which an effective radius of $\sim 1 \rm kpc$ has been reported (similar to the median value estimated for the HELM subsample at $z<1$). The attenuation inferred for CEERS-14821 refers to a compact system, whereas our radiative-transfer calculations integrate the emission over the full extent of the extracted regions, spanning $\sim 24$--$60$ physical kpc across the three galaxies.
The spatial scales involved therefore differ 
by more than an order of magnitude, partly reflecting the clumpy and still assembling nature of our simulated systems.

To explore the impact of spatial scale, panel b shows the cell-by-cell distribution of the visual attenuation. On these local scales, the two 
most massive galaxies (G22 and G5) reach, and in some regions exceed, the attenuation values characteristic of HELM systems. Among the 
transition-region galaxies, the local attenuation also increases with dust content, but even their most obscured regions remain below the HELM 
threshold. Interestingly, although one of the two metallicity estimates for CEERS-14821 is comparable to that of G818 and G3073 
(Fig.~\ref{fig:3scalingrelations}c), its inferred attenuation is substantially larger, placing it within the HELM selection region. 
Together with the high metallicities measured for the HELM galaxies, this suggests that these low-mass systems are significantly more 
chemically enriched, and consequently dustier, than simulated galaxies of comparable stellar mass.

\subsection{Synthetic JWST and ALMA observations}

\begin{figure*}
\centering

\includegraphics[width=\textwidth,height=8.0cm]{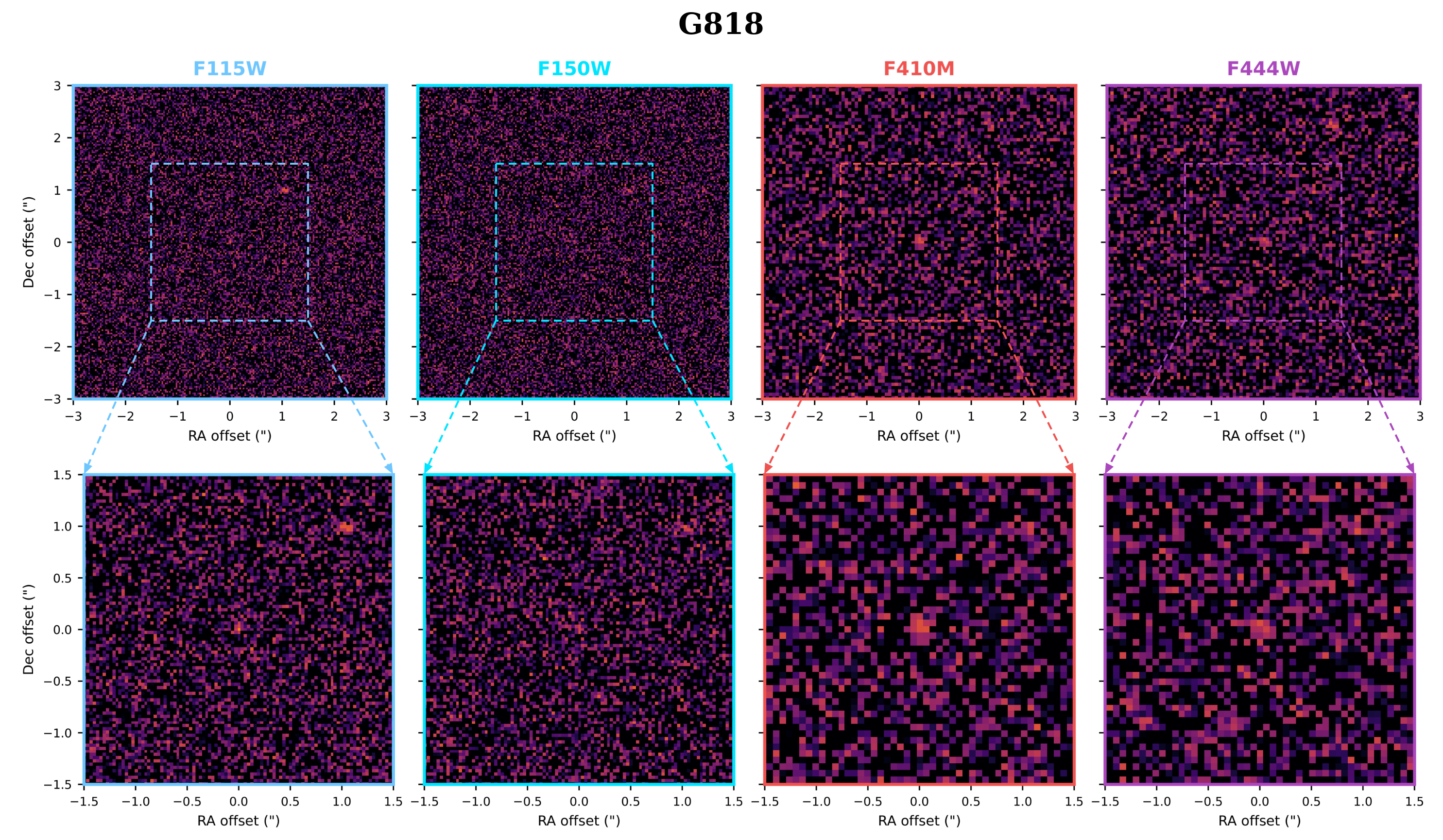}\\[-1mm]

\includegraphics[width=\textwidth,height=8.1cm]{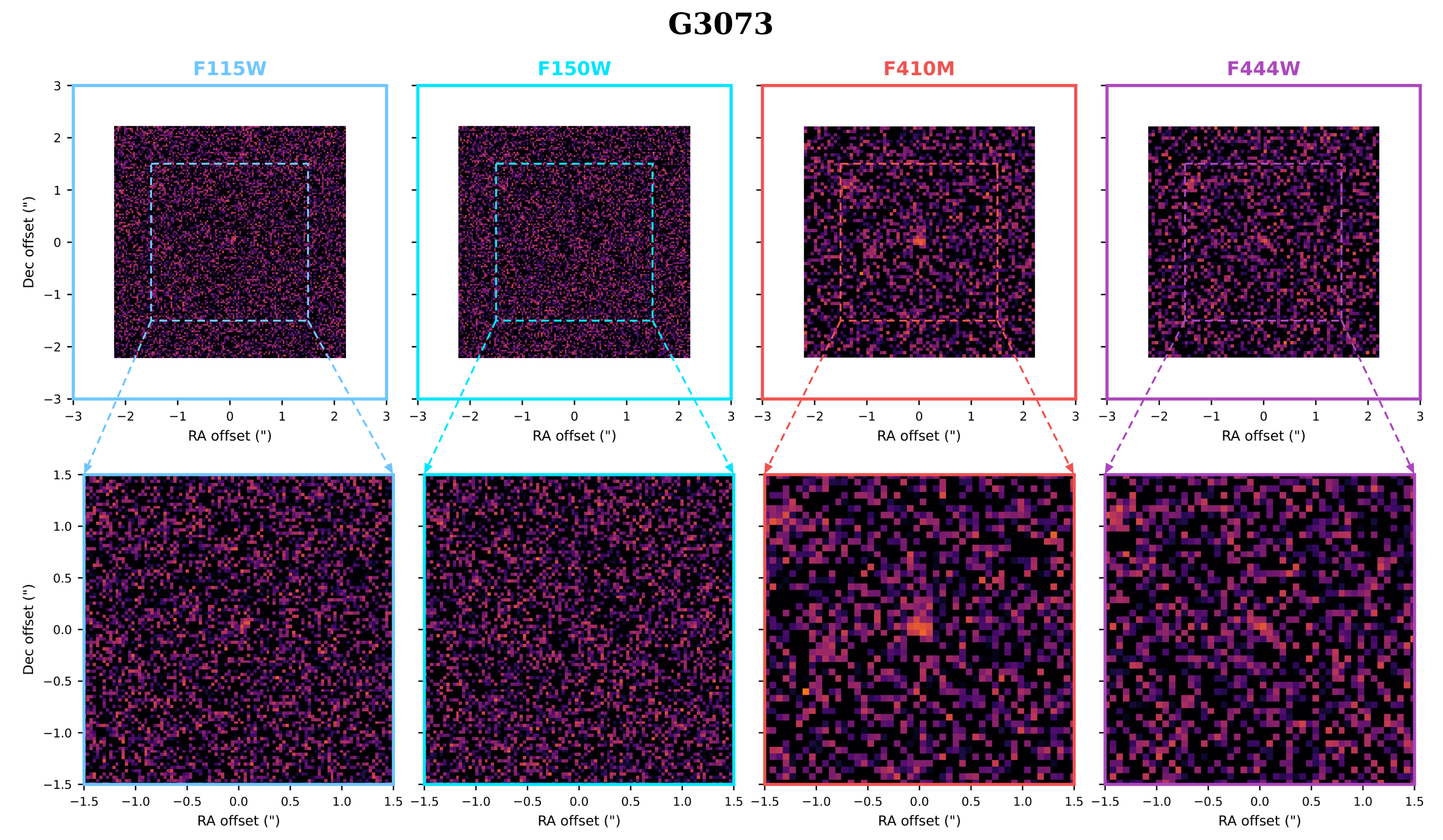}\\[-1mm]

\includegraphics[width=\textwidth,height=8.1cm]{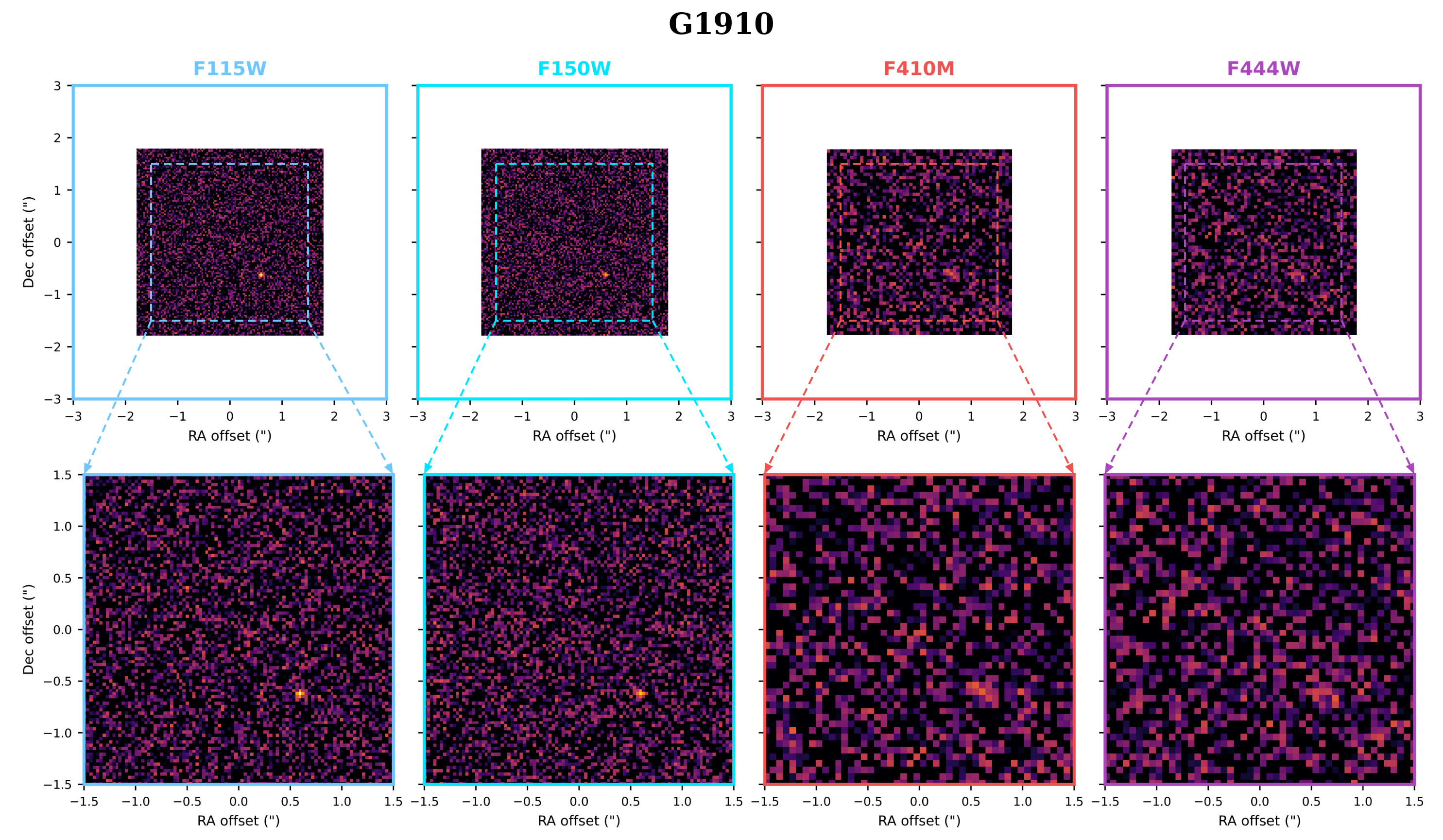}
\caption{Predicted JWST/NIRCam images of the selected galaxies in filters probing the rest-frame UV continuum (F115W, F150W), and the optical continuum (F410M, F444W). See the text for details of the procedure adopted to generate the mock observations.}
\label{fig:JWSTmock}
\end{figure*}

\begin{table*}

\centering
\small
\begin{tabular}{|l||cc|cc|cc|cc|cc|}
\hline
\multirow{2}{*}{} &
\multicolumn{2}{c|}{\textbf{F115W}} &
\multicolumn{2}{c|}{\textbf{F150W}} &
\multicolumn{2}{c|}{\textbf{F200W}} &
\multicolumn{2}{c|}{\textbf{F410M}} &
\multicolumn{2}{c|}{\textbf{F444W}} \\
\cline{2-11}
 & $t_{\rm needed} [\rm h]$ & $t_{\rm HELM} [\rm h]$ &
   $t_{\rm needed} [\rm h]$ & $t_{\rm HELM} [\rm h]$ &
   $t_{\rm needed} [\rm h]$ & $t_{\rm HELM} [\rm h]$ &
   $t_{\rm needed} [\rm h]$ & $t_{\rm HELM} [\rm h]$ &
   $t_{\rm needed} [\rm h]$ & $t_{\rm HELM} [\rm h]$ \\
\hline

\textbf{G818}
 & 0.9 & 1.6
 & 0.8 & 0.8
 & 0.8 & 0.8
 & 0.3 & 0.8
 & 0.5 & 0.8 \\
\hline

\textbf{G3073}
 & 9.9 & 1.6
 & 11.1 & 0.8
 & 1.1 & 0.8
 & 0.3 & 0.8
 & 0.8 & 0.8 \\
\hline

\textbf{G1910}
 & 235.3 & 1.6
 & 39.4 & 0.8
 & 6.4 & 0.8
 & 2.6 & 0.8
 & 1.0 & 0.8 \\
\hline

\textbf{G22}
 & 12.0 & 1.6
 & 4.3 & 0.8
 & 3.2 & 0.8
 & 0.1 & 0.8
 & 0.1 & 0.8 \\
\hline

\textbf{G5}
 & 96.1 & 1.6
 & 77.0 & 0.8
 & 13.2 & 0.8
 & 0.3 & 0.8
 & 0.6 & 0.8 \\
\hline

\end{tabular}
\caption{Exposure times required to achieve a $5 \sigma$ for the target galaxies in the different JWST/NIRCam filters, compared with the corresponding HELM integration times.}
\label{tab:SNRJWST}
\end{table*}


\begin{figure*}[]
   \centering
   \includegraphics[width=0.95\textwidth]{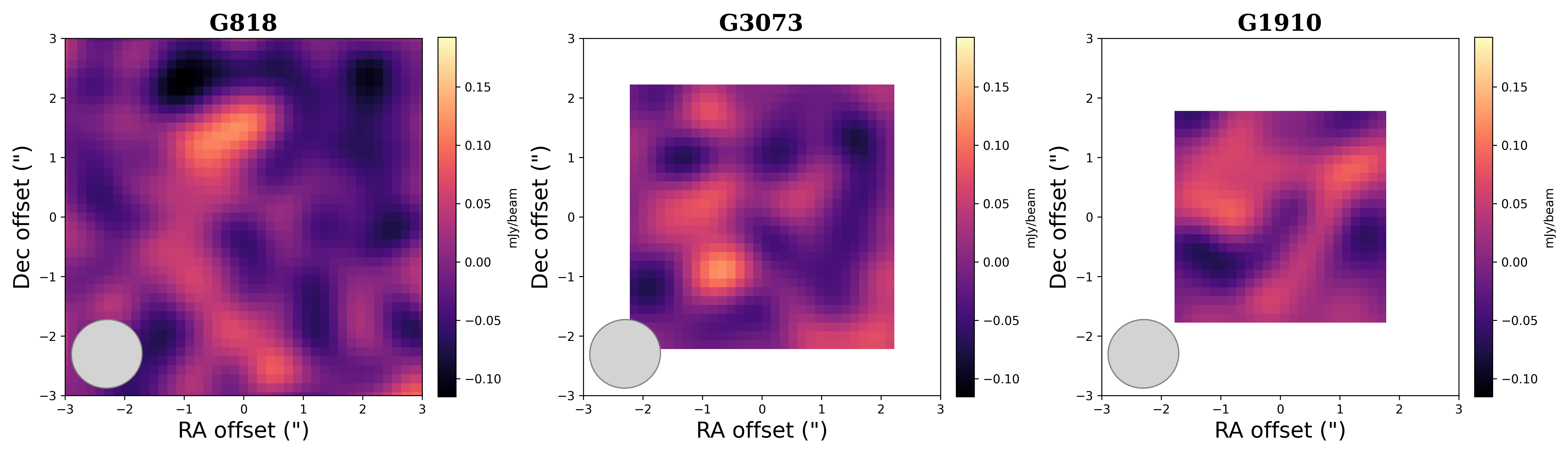}
   \caption{Mock ALMA Band 7 observations ($345 \, \mathrm{ GHz}$, observed frame) of the selected galaxies. Each map is the SKIRT surface brightness model convolved with the ALMA synthesised beam (filled grey circle in the bottom-left corner), resampled onto the ALMA pixel grid, and combined with Gaussian noise correlated on the beam scale. Beam ($1.19'' \times 1.15''$ FWHM) and rms ($38.5 \rm \mu Jy \cdot beam^{-1}$) are derived from an ALPINE-like ALMA GHOST simulation (20 min on source; see text).}  
\label{fig:ALMAmock}
\end{figure*}

To enable a direct comparison with observations, we use the frequency-dependent emissivity computed by SKIRT to generate synthetic surface brightness maps by redshifting the emission and, for JWST, integrating it over the wavelength range of each observational filter, while for ALMA we interpolate log-log between the two channels adjacent to the chosen band. We then construct mock JWST and ALMA observations by resampling the maps onto the corresponding detector grids, convolving them with Gaussian kernels to reproduce the instrumental resolution (filter-dependent PSFs for JWST/NIRCam; for ALMA, the synthesised beam obtained by applying the ALPINE set up to the SKIRT model with ALMA GHOST). Noise consistent with the expected sensitivities is added as a final step. \footnote{ Given that the ALMA synthesised beam ($1.19'' \times 1.15''$) and noise level ($38.5 \rm \mu Jy \cdot beam^{-1}$) depend only on the observational set-up (ALPINE-like: 20 min on source at $345 \, \rm GHz$, $43$ antennas, compact configuration) and not on the source, we adopt the same values for all five galaxies; this was also validated with independent GHOST runs for different exposure times and galaxies. }

\begin{table*}
\centering
\small
\begin{tabular}{|l||c|c|cc|}
\hline
\multirow{2}{*}{\textbf{}} &
\multirow{2}{*}{$\mathrm{S_{ peak}}\ [\mu{\rm Jy}]$} &
\multirow{2}{*}{SNR (20 min)} &
\multicolumn{2}{c|}{$t_{\mathrm{needed}}$} \\
\cline{4-5}
 & & & $3\sigma$ & $5\sigma$ \\
\hline

\textbf{G818}
 & 0.3 & 0.01
 & $\sim 5\times10^{4}$ h & $\sim 1.4\times10^{5}$ h \\
\hline

\textbf{G3073}
 & 1.9 & 0.05
 & $\sim$1300 h & $\sim$3600 h \\
\hline

\textbf{G1910}
 & 5.4 & 0.14
 & $\sim$150 h & $\sim$420 h \\
\hline

\textbf{G22}
 & 300 & 7.8
 & $\sim$3 min & $\sim$8 min \\
\hline

\textbf{G5}
 & 1300 & 34
 & $<1$ min & $<1$ min \\
\hline

\end{tabular}
\caption{Peak flux density, SNR achieved in 20 minutes of on-source time, and on-source time required to reach a $3\sigma$ and $5 \sigma$ detection with ALMA Band 7. Sensitivities were computed with ALMA GHOST (\url{https://alma-ghost.jb.man.ac.uk/}). }
\label{tab:SNRALMA}
\end{table*}

Since our modelling does not include nebular emission, we restrict the analysis to JWST/NIRCam filters that primarily trace the stellar continuum while minimising contamination from strong emission lines.
Specifically, we use the F115W and F150W bands to probe the UV continuum, and F410M and F444W to trace the optical continuum of galaxies at $z\sim 4.5$. This filter selection follows the observational strategy adopted for HELM galaxies, enabling a direct assessment of the detectability of our simulated galaxies under comparable observational conditions. For ALMA, we focus on Band 7, which probes the rest-frame dust continuum around $160 \, \rm \mu$m at $z = 4.5$. 

The resulting JWST/NIRCam and ALMA mock observations are shown in Figs. \ref{fig:JWSTmock} and \ref{fig:ALMAmock} for the three galaxies selected within the dust-scattered transition region. To facilitate a direct comparison between their intrinsic (Fig. \ref{fig:grids}) and observed morphologies, all images are displayed on a common spatial scale, chosen to encompass the full extent of the star-forming clumps while remaining consistent with the angular resolution of both observatories.
For the JWST/NIRCam images, we additionally show a zoom-in view of the central region to better resolve the dominant star-forming clump, which is used as the reference for estimating the required exposure times.

We estimate the exposure requirements separately for each facility. For JWST/NIRCam, we derive the integration time required for a $5 \, \sigma$ detection by rescaling the HELM exposure time according to the SNR achieved by each simulated galaxy under a HELM-like observational setup, adopting the standard $\mathrm{t \propto SNR^2}$  scaling. Fluxes are measured on the noisy mock images, within circular apertures centred on each galaxy, adopting a radius equal to the PSF FWHM for the lower-mass galaxies and 1.5 times larger for the more massive systems. The noise is Gaussian and uncorrelated between pixels, with a per-pixel rms derived from the $5 \sigma$ depths published for the CEERS/HELM observations \citep{Bisigello2023, bisigello2025helmsdeephighlyextincted}, and a single noise realisation is used for each galaxy and filter. The resulting exposure times are summarised in Table \ref{tab:SNRJWST}.

The JWST mock observations, together with the SNR estimates, suggest that the detectability of these galaxies depends strongly on the wavelength range being probed. For the dustiest transition-region galaxy, G1910, the central star-forming clump is heavily obscured in the bluest filters (F115W, F150W), requiring prohibitively long exposure times to recover the rest-frame UV continuum.

In contrast, the external clump remains detectable and dominates the observed emission at these wavelengths. At longer wavelengths, where dust attenuation is less severe, the central component becomes progressively more visible. G3073 can be detected in both of its main clumps across the full filter set, although with a marked wavelength dependence. Filters sampling the rest-frame optical continuum provide the most favourable detection, whereas the UV bands require substantially longer integrations, illustrating how multi-band imaging can recover the overall morphology despite the effects of dust attenuation. 
The least dusty system, G818, remains detectable primarily through its two dominant star-forming clumps, which contain most of its stellar mass and dust. Its central component requires exposure times comparable to, or shorter than, those adopted for the HELM galaxies.
The two more massive galaxies (G22, G5) follow the expected behaviour of chemically evolved systems: both remain easily detectable in the rest-frame optical, whereas the dustiest becomes progressively fainter towards shorter wavelengths, requiring unrealistically long integrations in the UV.

ALMA observations paint a different picture. At the depth of the ALPINE survey (20 min on source), we measure the SNR on the noise-free maps convolved with the synthesised beam, taking as noise the rms obtained from the GHOST simulation, and rescale the exposure times with the same $\mathrm{t \propto SNR^2}$ scaling adopted for JWST. The two more massive galaxies (G5, G22), selected to resemble ALPINE-like sources, are readily detected, reaching $\mathrm{SNR}>5$. In contrast, the dust continuum emission from the three low-mass transition region galaxies would require between $\sim 150 \rm h$ and $\sim 5\times10^{4}\rm h$  of on-source time even for a $3 \sigma$ detection (see Table \ref{tab:SNRALMA} for further details), and is therefore not observable with current ALMA, preventing their characterisation through continuum observations alone.

\section{Discussion} 
\label{sec:discussion}

\subsection{Is the local picture of dust evolution already in place at high redshift?}
\label{subsec:disc-picture}

The first question motivating this work is whether the physical picture of dust evolution established in nearby galaxies already applies within the first billion years of cosmic history.

Our results suggest that the answer is affirmative. The comparison between the total dust masses predicted by the simulations and the contribution from stellar sources alone (empty black symbols in Fig.~\ref{fig:3scalingrelations}) shows that stellar dust production can reproduce only the low-dust branch of the dust-to-stellar mass relation. Grain growth is therefore required at $z\sim4.5$ to explain both the change in slope and the large dispersion of the relation. In this context, such an S-shaped transition has been recently modeled using a phenomenological hybrid approach to demonstrate how a shift in the dominant origin of dust can naturally explain the extremely low dust attenuation and the unexpected excess of UV-bright galaxies at $z \geq 9$ \citep{Burgarella2026}.

This picture is fully consistent with theoretical models of dust evolution developed for nearby galaxies, in which the transition from stellar dust production to grain growth naturally gives rise to the characteristic S-shaped dust-to-stellar mass and dustiness-metallicity relations \citep[e.g.][]{Asanoetal2013,sorgentidipolvere,Galliano_anearbyperspectiveondustevolution}. More recently, \citet{Di_Cesare_2023} showed that cosmological simulations already reproduce these scaling relations at high redshift, while observational studies with JWST are beginning to reveal the same behaviour in galaxies at $z\gtrsim4$ \citep{burgarella2025ceersforgingdustgrains}. Our results provide a physical interpretation of this emerging picture by directly isolating the stellar contribution to the dust budget and demonstrating that grain growth is already the dominant mechanism regulating dust masses in the transition region. Our conclusions are also broadly consistent with recent theoretical work showing that efficient grain growth can already be established at very early cosmic epochs once the evolution of the grain-size distribution is taken into account \citep{toyouchi2026}. Although our simulations do not explicitly follow grain-size evolution, both studies point toward the same conclusion: the conditions required for rapid dust growth can be established well before galaxies reach the ISM conditions (density, temperature, and metallicity) typical of nearby systems.

At the same time, our simulations indicate an important difference with respect to the local Universe. While nearby galaxies undergo the transition to efficient grain growth around $Z \sim0.2\,Z_\odot$ \citep{Galliano_anearbyperspectiveondustevolution}, our galaxies exhibit the change in slope of the dustiness-metallicity relation at significantly lower metallicity ($Z\sim0.04\,Z_\odot$). This trend is also hinted at observationally: JWST/NIRspec spectroscopy of galaxies at $4.0 < z \leq 11.4$ places the empirical transition at $Z \sim 0.1 \, Z_\odot$, corresponding to $12+\log(\text{O/H}) \approx 7.60$ \citep{burgarella2025ceersforgingdustgrains}. This value, while consistent with model predictions, sits toward the lower end of the expected range following the same direction found in our simulations. Within the framework proposed by \citet{Asanoetal2013}, this behaviour is naturally explained by the higher gas densities typical of early galaxies, which shorten grain-growth timescales and allow efficient accretion onto pre-existing grains despite the lower metal abundance.

Finally, our simulations indicate that metallicity alone cannot fully explain the diversity of dust masses observed at high redshift, suggesting that additional physical parameters regulate grain growth. 

\subsection{What drives the scatter in dust content?}
\label{subsec:disc-scatter}

Since grain growth is expected to depend primarily on the chemical enrichment history of galaxies \citep{Asanoetal2013,Galliano_anearbyperspectiveondustevolution}, we first examine their star formation histories (see Fig. \ref{fig:SFH_complete}). As expected, progressively longer star formation histories lead to higher metallicities and larger reservoirs of metals available for grain growth. G818, for example, has the most recent onset of star formation, lacks solar-metallicity gas, and is the only system in which stellar dust still dominates the total dust budget.

Yet, metallicity alone cannot fully explain the observed diversity. Galaxies with similar metallicities still exhibit significantly different dust contents, indicating that another parameter must regulate the efficiency of grain growth. We therefore explored whether this could be linked to the large-scale environment. Although recent observations have suggested that galaxy interactions and dense environments may favour rapid dust enrichment \citep{bisigello2025helmsdeephighlyextincted}, all three galaxies occupy similar nodes of the cosmic web, providing no evidence that their large-scale environment is the primary driver of the observed scatter.

The clearest discriminator instead emerges from their internal structure. All three galaxies are clumpy systems still assembling at these redshifts, consistent with recent JWST observations of low-mass galaxies \citep[e.g.][]{Faisst_massmetallicity}. However, their degree of assembly closely follows their dust content: progressively more assembled galaxies exhibit denser gas distributions, higher metallicities, and increasingly efficient grain growth. Rather than being the physical cause of dust growth, morphology therefore acts as an observable tracer of the evolutionary stage of the galaxy and of the ISM conditions that regulate grain growth.

Our results therefore extend the picture proposed by \citet{Di_Cesare_2023} by suggesting that the large scatter around the dust scaling relations is itself an evolutionary effect, reflecting different stages of galaxy assembly rather than distinct classes of galaxies.

\subsection{Can current observations reveal this evolutionary sequence?}
\label{subsec:disc-observations}

If morphology traces the evolutionary stage of galaxies, the next question is whether this evolutionary sequence can be recognised observationally.

Our synthetic SEDs suggest that the answer depends critically on the adopted dust tracer. Although the overall spectral properties of our galaxies are broadly consistent with current observations at similar redshifts \citep{Fudamoto,Gruppioni,Bouwens}, the UV continuum slope alone does not provide a reliable measure of dust content. In particular, G3073 and G1910 exhibit comparable dust masses but markedly different UV slopes because of the different spatial distributions of young stars and dusty gas. This confirms that the UV slope is affected by degeneracies involving both star-dust geometry and stellar population age, consistent with previous observational and theoretical studies \citep[e.g.][]{Salim2020,rodighiero2026empiricalderivationaveragedust}. In our simulations, V-band attenuation and the infrared excess (IRX) provide more robust diagnostics of dust content.

Current JWST observations already offer an important benchmark for these predictions. The highly obscured low-mass galaxies identified by the HELM survey (\citealt{ bisigello2025helmsdeephighlyextincted}; see also \citealt{Bisigello_2023, Gandolfi2026}), and in particular CEERS-14821 \citep{Bisigello_2025}, demonstrate that substantial dust attenuation can already be present in galaxies of comparable stellar mass. However, none of our simulated systems reaches the extreme attenuation inferred for these objects (Fig.~\ref{fig:AVcomparisons}), even when considering the full range of smaller-scale values predicted by the radiative-transfer calculations. These galaxies may therefore represent particularly dust-rich systems, possibly occupying the upper envelope of the dust distribution at fixed stellar mass.

To assess whether the predicted structural differences are directly observable, we generated mock JWST and ALMA observations. Multi-band JWST imaging successfully recovers the dominant stellar clumps, allowing the differences in clump number and distribution across galaxies to be distinguished, although the faintest clumps remain difficult to reconstruct in the least evolved systems. In contrast, the predicted dust continuum emission lies well below the sensitivity limits of current ALMA continuum surveys, with fluxes 2--3 orders of magnitude lower than those measured for galaxies of comparable stellar mass at similar redshifts \citep{Valentinoetal_ALMASED}. While JWST is therefore already capable of revealing differences in the clumpiness of galaxies at different stages of assembly through their resolved stellar emission, direct measurements of their dust reservoirs remain largely restricted to the most dust-rich systems. 

Future ALMA continuum surveys with substantially improved sensitivity will therefore be essential to establish whether the evolutionary sequence proposed here is representative of the broader population of low-mass galaxies in the early Universe.

\subsection{Limitations and future prospects}
\label{subsec:disc-caveats}

Our interpretation of dust evolution at high redshift is necessarily subject to a number of modelling assumptions that should be kept in mind when interpreting the results.

The current dust model implemented in \texttt{ dustyGadget} follows grain growth through a simplified prescription based on metal accretion onto pre-existing grains and assumes a fixed grain size, without explicitly tracking the evolution of different grain species or grain-size distributions. Although these simplifications may affect the detailed dust properties, the simulations successfully reproduce a wide range of observed scaling relations of high-redshift galaxies \citep{Di_Cesare_2023}, providing confidence that they capture the dominant physical processes regulating dust evolution.
These details, however, are expected to become critical for a future computation of the dust optical properties.
Future developments including updated stellar dust yields and a realistic grain-size distribution will therefore allow these conclusions to be tested with a more complete dust model.

Another limitation concerns the numerical resolution. The dense, cold phases of the interstellar medium where grain growth is expected to be most efficient are not resolved directly but described through sub-grid prescriptions, as is common in current cosmological simulations with dust physics (e.g. \citealt{aoyama2017,mckinnon2017,li2019,vogelsberger2019,choban2022,lewis2023,dubois2024,esmerian2024}). Higher-resolution simulations capable of resolving the multiphase ISM will be essential to quantify more accurately the efficiency and spatial distribution of grain growth \citep{Trayford2026, Vijayan2026}.
A related caveat to the mass resolution concerns the identification of the systems we refer to as \textit{galaxies}, which also depends partly on the halo-finding method. G818 provides a striking example of how multiple distinct baryonic components can be identified within a single halo. Higher-resolution simulations may help clarify the nature of these components, while observationally it remains an open question whether they would be identified as separate galaxies or as parts of a single system. Clumpy systems with multiple, often interacting components are, however, observed at similar redshifts \citep{Faisst_massmetallicity}, supporting the physical plausibility of such configurations.

Our observational predictions should likewise be regarded as indicative. The mock JWST and ALMA observations are based on continuum emission only, assume idealised noise properties, and consider a single viewing angle, while real observations are affected by correlated noise, line contamination, and orientation-dependent attenuation. These simplifications introduce additional uncertainties in the predicted detectability of individual systems, but they are not expected to alter the qualitative conclusion that JWST is considerably more sensitive than current ALMA continuum surveys to the low-mass galaxies considered here.

Finally, our analysis focuses on five representative galaxies, with particular attention devoted to three systems spanning the transition region of the dust-to-stellar mass relation. Although this targeted approach allows us to identify the physical mechanisms responsible for the observed diversity in dust content, establishing how generally this evolutionary sequence applies will require analyses of substantially larger simulated samples together with deeper ALMA continuum observations capable of probing the dust reservoirs of typical low-mass galaxies at high redshift.

\section{Conclusions}
\label{sec:conclusions}

In this work we investigated the origin of the scatter in the dust-to-stellar mass relation at $z \simeq 4.5$ using cosmological hydrodynamical simulations, radiative transfer calculations, and synthetic JWST and ALMA observations. By selecting galaxies spanning the transition region of the dust-to-stellar mass relation ($\mathrm{log(M_{\star}/M_{\odot}) \sim 8.5 - 9.0}$), we explored the connection between dust enrichment and galaxy assembly, and its observational implications. Our main conclusions are as follows.

\begin{enumerate}
\item The physical picture of dust evolution established in nearby galaxies is already in place at $z \simeq 4.5$. The simulated galaxies reproduce the characteristic S-shaped dust-to-stellar mass and dustiness--metallicity relations, indicating the transition from stellar dust production to grain growth. Compared to the local Universe, this transition appears to occur at lower metallicities, consistent with the higher gas densities typical of early galaxies.

\item Stellar dust production alone cannot account for either the dust masses or the large scatter observed among galaxies with similar stellar masses. Comparison with our semi-analytical model demonstrates that grain growth is already the dominant mechanism regulating the dust budget of galaxies in the transition region, while stellar sources primarily determine the initial dust reservoir from which grain growth proceeds.

\item The observed scatter is primarily linked to the evolutionary stage of galaxy assembly. Although the selected galaxies have similar stellar masses, they exhibit markedly different star formation histories, morphologies, metallicities, and dust contents. These properties consistently suggest an evolutionary sequence in which progressively more assembled galaxies develop denser and more chemically enriched ISM conditions, leading to increasingly efficient grain growth. Morphology is therefore best interpreted as an observable tracer of galaxy assembly, rather than the physical driver of dust evolution.

\item The synthetic observables highlight both the opportunities and limitations of current facilities. UV continuum slopes alone do not provide a reliable measure of dust content because they depend strongly on the relative geometry of stars and dust. Conversely, quantities such as $A_V$, IRX, and the infrared luminosity correlate more closely with the total dust mass in the ISM. Mock observations further show that JWST imaging can identify galaxies in the transition regime and reveal wavelength-dependent structural variations, whereas ALMA continuum observations with sensitivities comparable to ALPINE remain largely limited to the most dust-rich systems.
\end{enumerate}

Our results suggest that the large scatter around the high-redshift dust scaling relations is not simply stochastic, but reflects different stages of galaxy assembly. Rather than representing fundamentally distinct populations, galaxies populating the transition region of the dust-to-stellar mass relation appear to be different snapshots of a common evolutionary pathway, in which progressively denser and more chemically enriched ISM conditions enable the onset of efficient grain growth. This interpretation extends the picture emerging from recent cosmological simulations \citep{Di_Cesare_2023} by showing that the scatter itself carries information on the evolutionary state of galaxies.

While current JWST observations already provide key diagnostics of stellar populations, metallicity, and galaxy structure, direct measurements of the corresponding dust reservoirs remain beyond the reach of current ALMA continuum surveys for typical low-mass galaxies. Future improvements in continuum sensitivity, particularly those enabled by the ALMA2030 Wideband Sensitivity Upgrade \citep{Carpenter2023}, will therefore be essential to test this evolutionary scenario across representative samples in the early Universe.

\begin{acknowledgements}
We acknowledge support from the MUR projects FIS-2024-01621 DAWN and PRIN 2022CB3PJ3-FLAGS, and from the EU-Recovery Fund PNRR. CDC acknowledges support from the European Union (ERC, AGENTS, 101076224).
\end{acknowledgements}

\bibliographystyle{aa} 
\bibliography{bibliografia} 

\begin{appendix}

\onecolumn

\section{Evolution of the stellar contribution to dust up to $z \approx 4.5$}
\label{sec:appendix_stardustevolution}

\begin{figure}[ht!]
   \centering
   \includegraphics[width=0.65\textwidth]{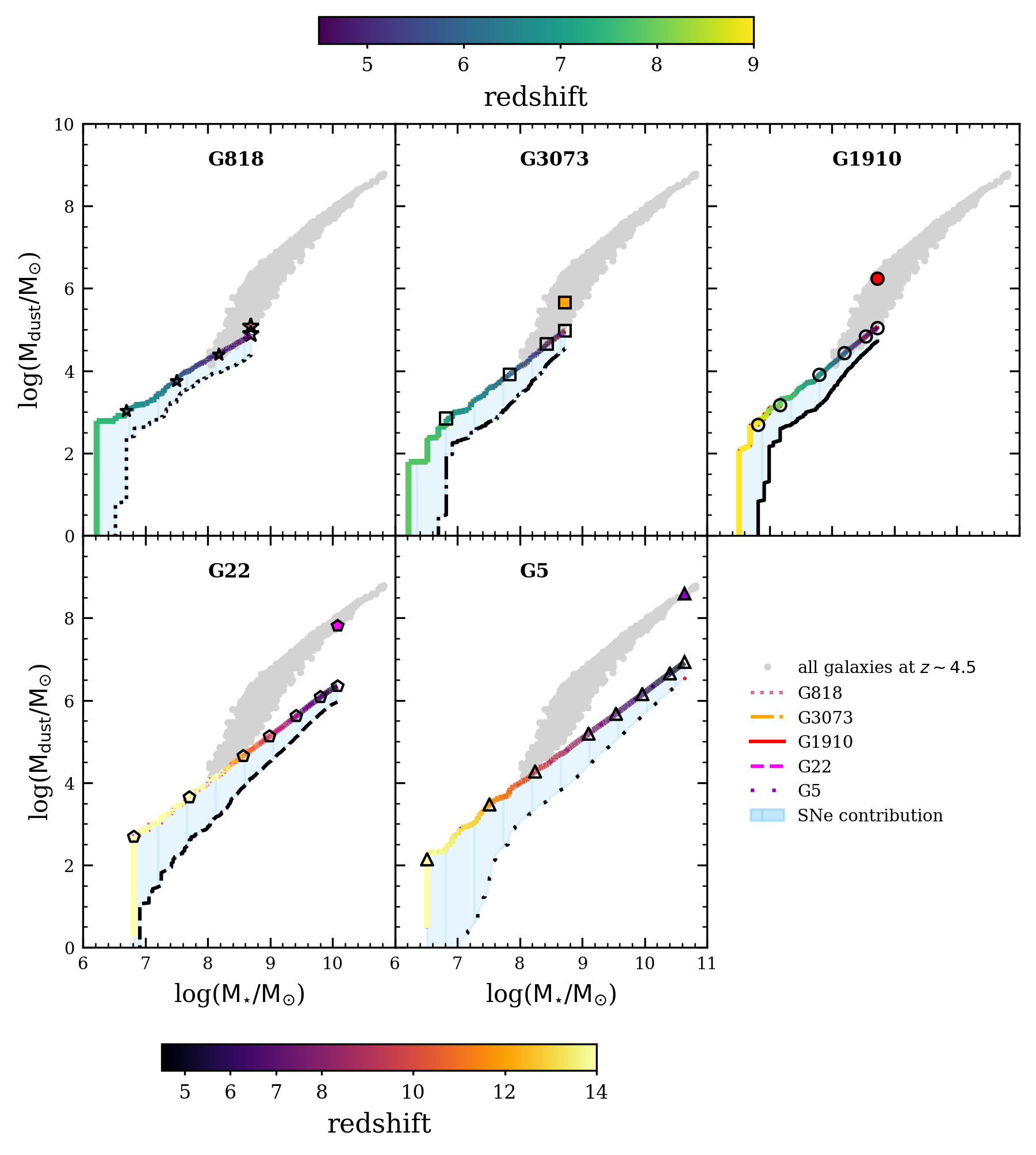}
   \caption{Evolution of the total stellar dust mass predicted by the semi-analytical model, colour-coded by redshift, overplotted on the dust-to-stellar mass relation at the snapshot time, where each point represents a simulated galaxy. Symbols along each track mark the redshifts highlighted in the colour bar, up to the final snapshot value. Black curves show the AGB-only contribution to the total stellar dust mass, following the galaxy-specific line styles adopted in Fig. \ref{fig:SFH_complete}, while the shaded regions indicate the additional contribution from SNe. Interpreted in an evolutionary context, the tracks consistently trace the expected stellar-dust-dominated range, departing from it at redshifts that depend on the specific SFH, with the more massive galaxies already departing at $z \lesssim 10$. }  
\label{fig_app:stardustevolution}
\end{figure}

\clearpage
\section{Predictions for the most massive galaxies in our sample}
\label{sec:appendix_predictionsmassivegal}

\begin{figure}[ht!]
\centering

\includegraphics[width=0.6 \textwidth]{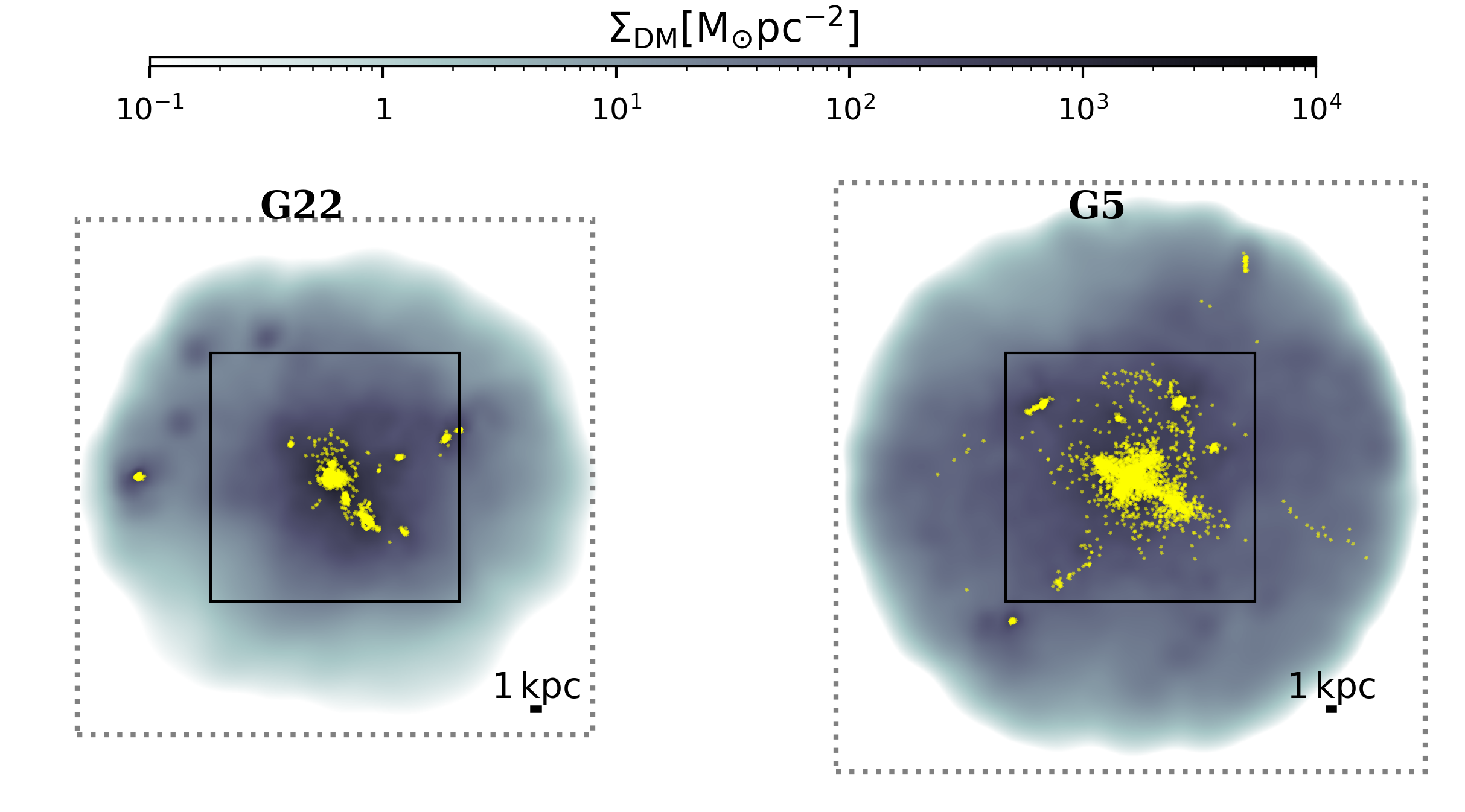}\\[-1mm]
\includegraphics[width=0.6\textwidth]{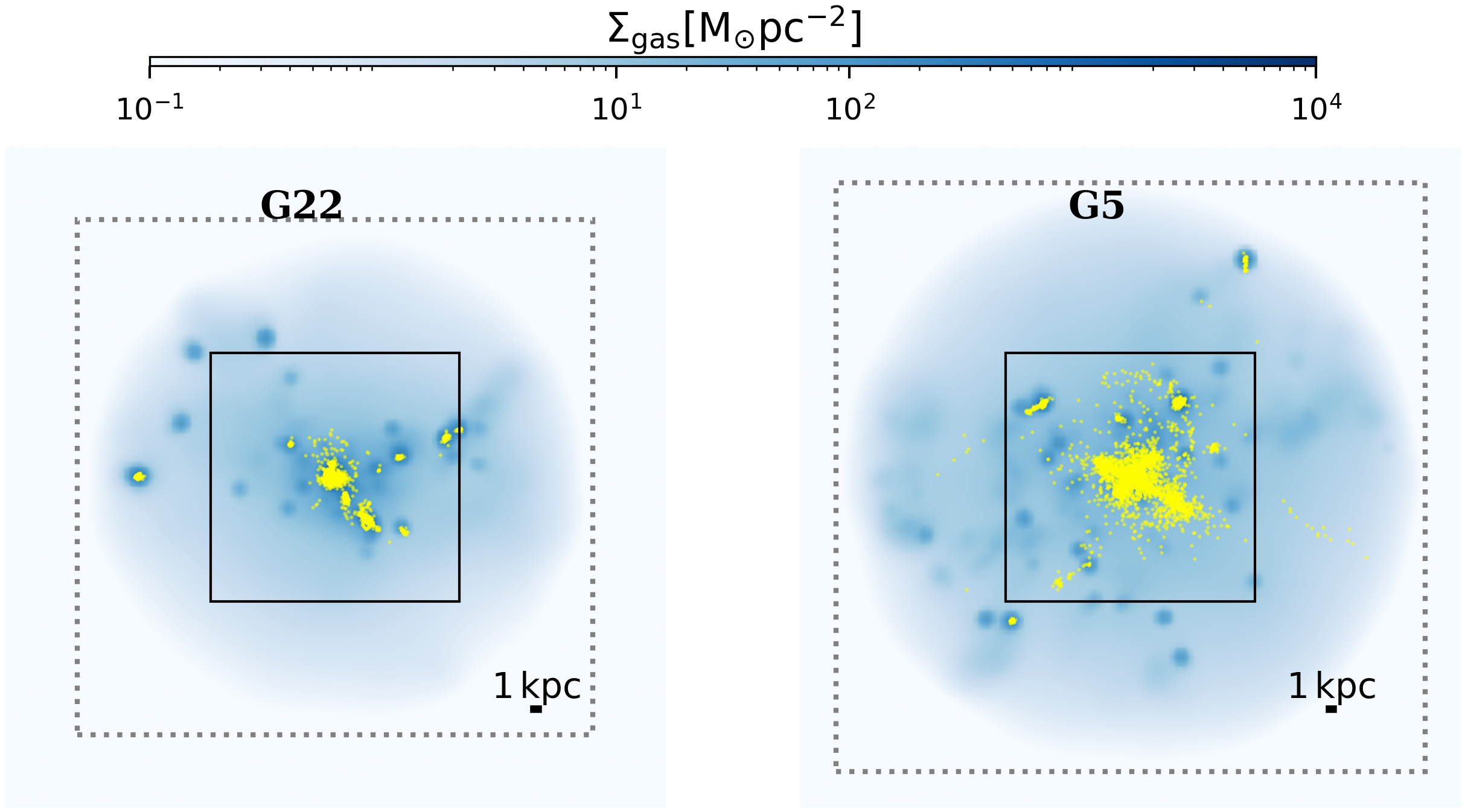}\\[-1mm]
\includegraphics[width=0.6 \textwidth]{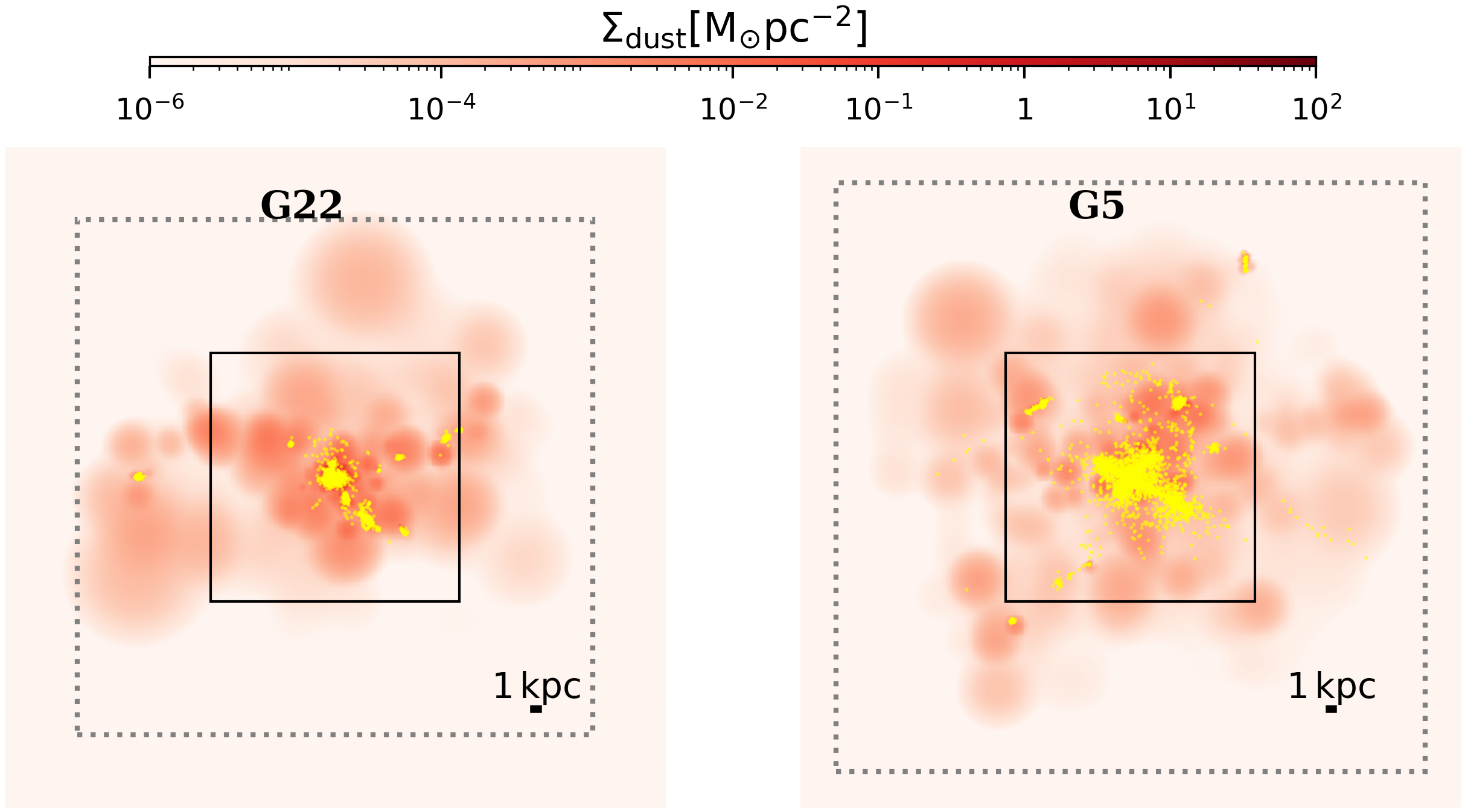}

\caption{The top, middle, and bottom panels show the dark matter, gas, and dust surface density maps, respectively, for the massive galaxies selected from the dust-to-stellar mass relation. The description of the plotted quantities is the same as in Fig. \ref{fig:grids}. Unlike the central zoom-in shown for the three galaxies discussed in the main text, here the maps span a larger field of view that extends beyond the virial radius, while the underlying particle selection is still restricted to $R_{\rm vir}$: this captures the more extended stellar populations of these more massive, evolved systems while excluding contamination from nearby haloes.}
\label{fig_app:grids_massivegal}
\end{figure}

\begin{figure}[ht!]
   \centering
   \includegraphics[width=0.45\textwidth]{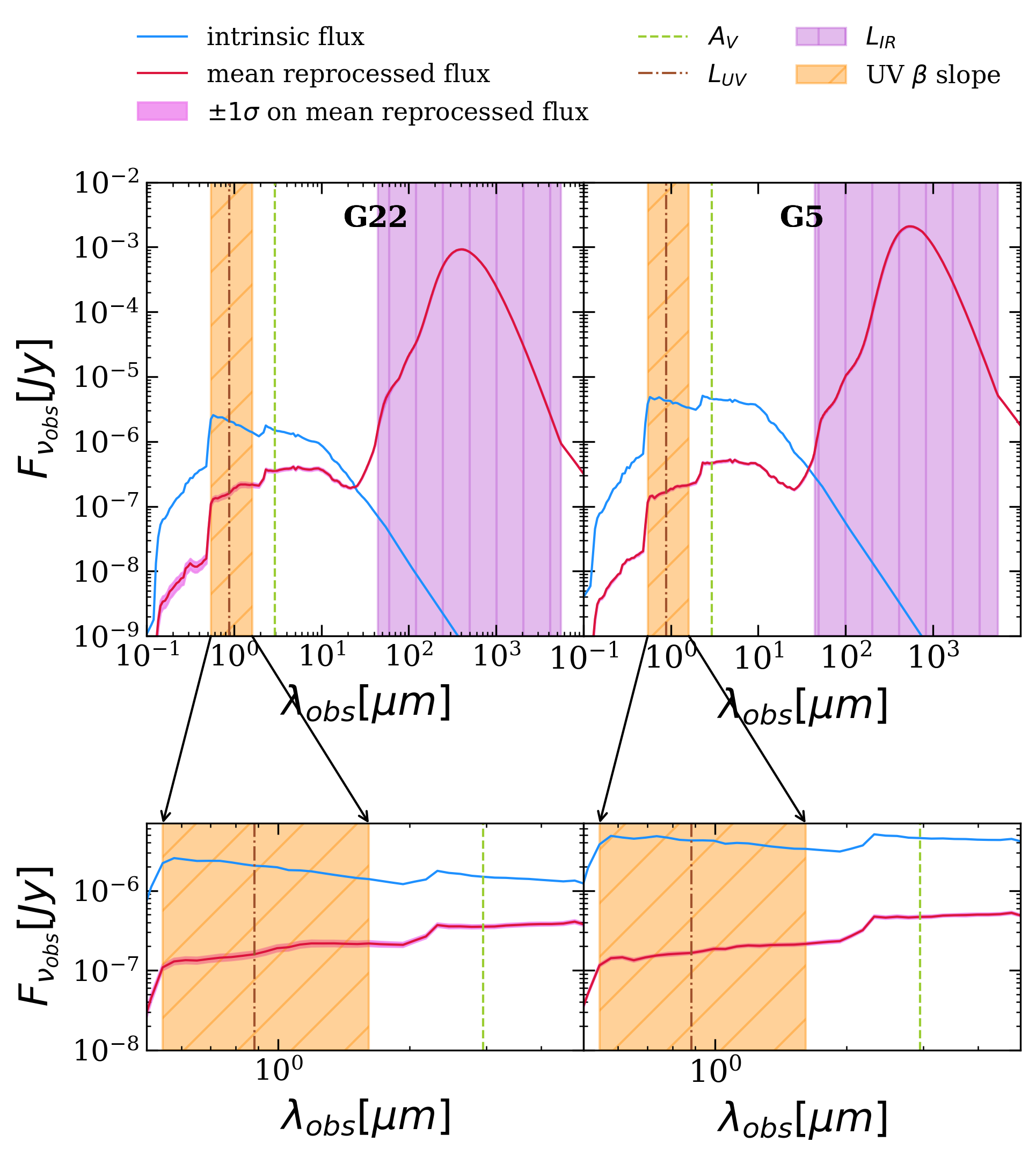}
   \caption{Synthetic SEDs of the most massive galaxies in our selected sample at $z \sim 4.5$. The description of the plotted quantities is the same as in Fig. \ref{fig:spectra_focusonspectralanalyses}. }  
\label{fig_app:spectra_massivegal}
\end{figure}

\begin{figure}[ht!]
\centering
\includegraphics[width=\textwidth,height=5.5cm]{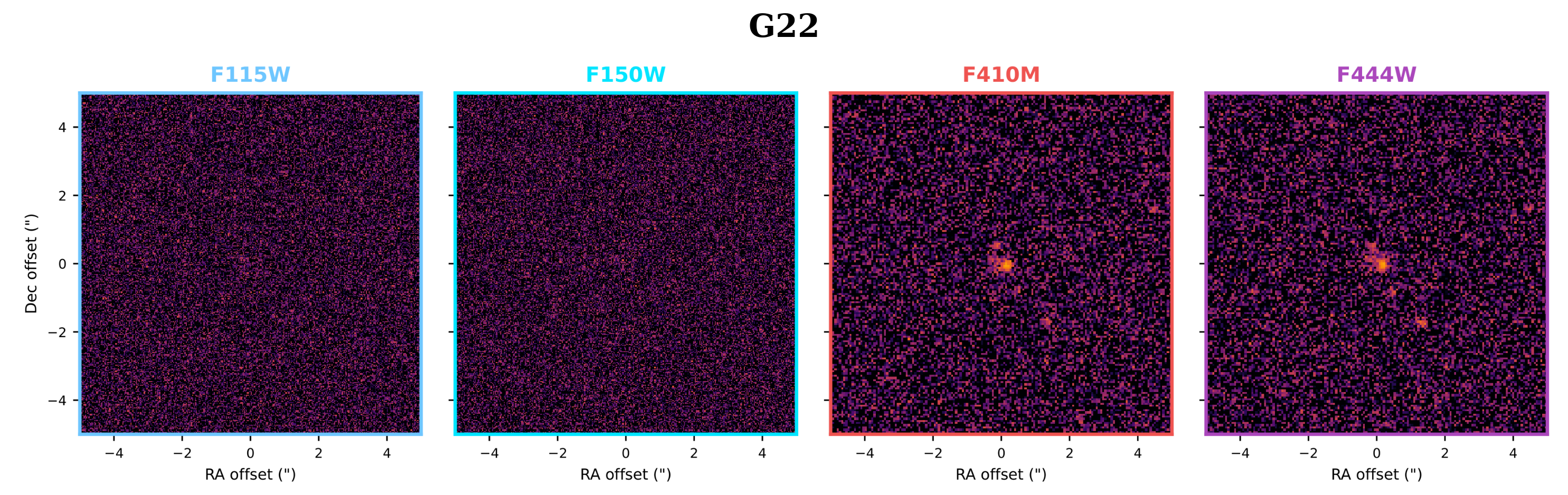}\\[-1mm]
\includegraphics[width=\textwidth,height=5.5cm]{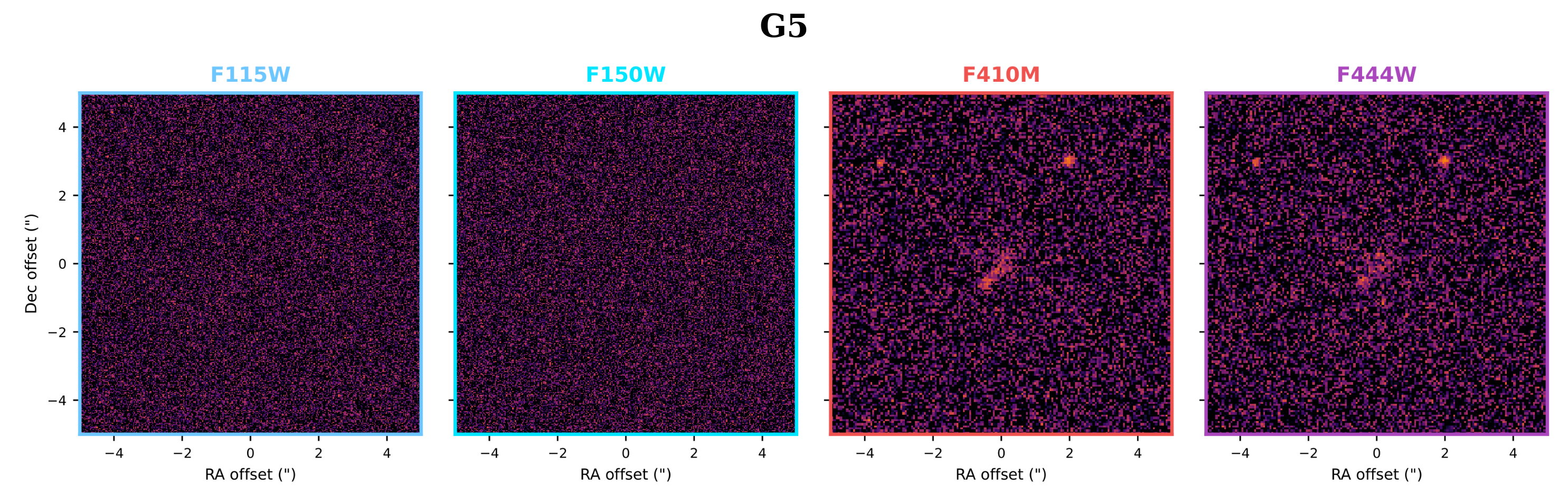}\\[-1mm]

\caption{Predicted JWST/NIRCam images of the massive galaxies, as in Fig.\ref{fig:JWSTmock}. Here we directly show the zoom-in on the central region encompassing most of the stellar populations, indicated by the black square in the resolved maps of Fig. \ref{fig_app:grids_massivegal}.}
\label{fig:JWSTmock_massivegal_appendix}
\end{figure}

\begin{figure}[ht!]
   \centering
   \includegraphics[width=0.7\textwidth]{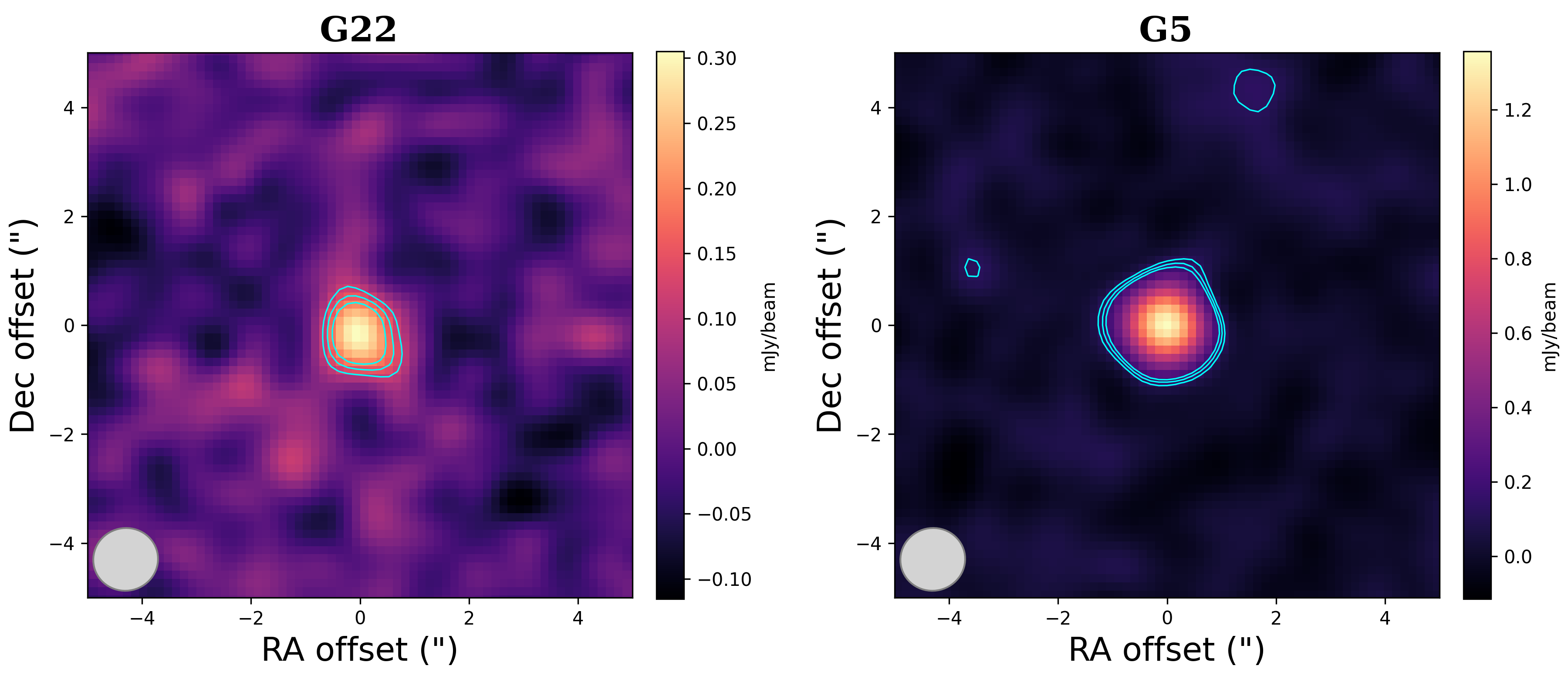}
   \caption{Predicted mock ALMA Band 7 observations, as in Fig. \ref{fig:ALMAmock}. The additional cyan contours mark the $3$, $4$ and $5 \sigma$ levels, where $\sigma$ is the rms noise.  }  
\label{fig_app:ALMAmock_massivegal}
\end{figure}


\end{appendix}

\end{document}